\documentclass[pdflatex,sn-mathphys-num]{sn-jnl}

\usepackage{graphicx}
\usepackage{multirow}
\usepackage{amsmath,amssymb,amsfonts}
\usepackage{booktabs}
\usepackage{subcaption}
\usepackage{float}
\usepackage{dblfloatfix}  
\usepackage{xcolor}  

\title[Redox behaviour of Fe impurities in BaTiO$_3$]{Redox behaviour of Fe impurities in BaTiO$_3$ based on many-body calculations}

\author[1]{\fnm{Zhiyuan} \sur{Li}}
\equalcont{These authors contributed equally to this work.}

\author[1]{\fnm{Hamza} \sur{Zerdoumi}}
\equalcont{These authors contributed equally to this work.}

\author[1]{\fnm{Hao} \sur{Wang}}

\author[1]{\fnm{Ruiwen} \sur{Xie}}

\author*[1]{\fnm{Hongbin} \sur{Zhang}}\email{hongbin.zhang@tu-darmstadt.de}

\affil[1]{\orgdiv{Institute of Materials Science},
           \orgname{Technische Universität Darmstadt},
           \orgaddress{\city{Darmstadt}, \postcode{64287}, \country{Germany}}}

\abstract{Based on detailed electronic structure and spectroscopy obtained using DFT-based many-body techniques, the redox behavior of Fe impurities in BaTiO$_3$ is investigated. It is observed that Fe impurities exhibit a mixed valence nature, comprising mostly  Fe$^{2+}$ ($3d^6$) and  Fe$^{3+}$ ($3d^5$) configurations, and such configurations can be tuned via oxygen vacancies which favor  Fe$^{2+}$. The origin of such a redox behavior can be attributed to the charge transfer caused by shifting of the $d_{3z^2-r^2}$ orbitals. Furthermore, x-ray photoemission spectroscopy is evaluated by solving the Wannier function-derived local atomic Hamiltonian using the crystal field multiplet approach, with good agreements with recent experimental measurements.}

\keywords{BaTiO$_3$, doping, DFT+DMFT, redox behavior, crystal field}

\begin{document}

\maketitle

\section{Introduction}\label{sec:intro}

Perovskite oxides, particularly barium titanate (BaTiO$_3$, BTO), constitute a cornerstone of modern electroceramics owing to their exceptional dielectric, ferroelectric, and piezoelectric properties \cite{cohen1992origin, scott2007applications}. Chemical substitution at the B-site offers a versatile route to tailor these functionalities, where different transition metal dopants give rise to remarkably distinct structural and electronic responses\cite{dietl2010ten, spaldin2005renaissance, nakayama2001theoretical, nossa2015effects}. While Mn, Co, Ni, and Zn substitutions in BTO primarily modify the sequence of temperature-driven transitions among conventional perovskite polymorphs \cite{keith2004synthesis}, Fe substitution is unique in stabilizing the high-temperature hexagonal phase ($P6_3/mmc$) at room temperature above a critical concentration of around 4\% \cite{li2026origin, nguyen2011tetragonal, rani2016structural, dang2011structural}. Additionally, Fe doping introduces localized magnetic moments into the ferroelectric host, giving rise to multiferroic functionality and magnetoelectric coupling \cite{xu2009room, mangalam2009multiferroic, ray2008high}. In order to elucidate such doping-induced structural transitions, the magnetic properties, and the optoelectronic properties\cite{noguchi2020successive}, it is pivotal to quantify the redox behavior of Fe impurities, as well as its interplay with charge-compensating oxygen vacancies ($\text{V}_{\rm O}$s), which is intimately tied to their local valence and spin state\cite{rani2016structural}.

However, despite extensive experimental studies, the redox behavior of Fe impurities in BTO remains an unresolved issue beyond simple ionic expectations, such as nominal Fe$^{4+}$ substitution for Ti$^{4+}$ \cite{adeagbo2019theoretical, mikulska2015x,jartych2018mossbauer, rani2016structural, noguchi2020successive}. Specifically, X-ray absorption spectroscopy (XAS) at the Fe $L$- and $K$-edges has consistently reported that the Fe impurities in BTO adopt a dominant Fe$^{3+}$ valence state \cite{mikulska2015x, chikada2010analysis, padchasri2021crystal}. Similarly, M\"{o}ssbauer isomer shifts and quadrupole splittings confirm that the Fe impurities exclusively exhibit a Fe$^{3+}$ valence state across distinct local environments\cite{jartych2018mossbauer}. Electron paramagnetic resonance (EPR) has also successfully identified specific paramagnetic Fe$^{3+}$ centers, including isolated impurities and Fe-oxygen vacancy (V$_\mathrm{O}$) pairs \cite{bottcher2018incorporation, langhammer2020incorporation}. However, these measurements cannot exclude the coexistence of the typically EPR-silent Fe$^{2+}$ species. Furthermore, X-ray photoelectron spectroscopy (XPS) core-level binding energies and satellite structures provide direct evidence for the coexistence of mixed Fe$^{3+}$ and Fe$^{2+}$ valence states \cite{chaoudhary2026direct, noguchi2020successive}. Macroscopically, X-ray diffraction (XRD) measurements corroborate such deviation, as observed lattice expansions point to the substitution of the smaller host Ti$^{4+}$ ($0.605$~\AA) by larger high-spin Fe$^{3+}$ ($0.645$~\AA) and Fe$^{2+}$ ($0.780$~\AA) \cite{rani2016structural}. Conversely, based on XAS and electrical measurements, the presence of Fe$^{4+}$ can only be inferred after treatments under strongly oxidizing conditions, demonstrating that the actual valence state is highly sensitive to the processing history\cite{maso2006electrical, nguyen2011tetragonal}. While charge-compensating $\mathrm{V_O}$s is conventionally invoked to explain these acceptor-like Fe$^{3+}$ or Fe$^{2+}$ states \cite{jartych2018mossbauer, ganegoda2021role, noguchi2020successive}, the available experimental observations alone cannot fully settle the physical origin of this complex redox response, demanding further theoretical treatment.

Theoretical resolution of this puzzle is challenged by the strongly correlated nature of the Fe $3d$ impurity electrons \cite{anderson1961localized, hubbard1963electron}. While previous first-principles studies using DFT and its U extention have successfully identified Fe-derived donor and acceptor levels within the BTO bandgap and their role in enhancing visible-light photoresponse \cite{noguchi2020successive,upadhyay2011enhanced}, these static mean-field descriptions, typically restricted to a single Slater determinant, may neglect the dynamic quantum fluctuations among various atomic multiplets. Recent DFT+DMFT studies on point defects in complex oxides highlight the necessity of going beyond these static limits: for Mn-substituted BaTiO$_3$, the ground state is revealed as a quantum superposition of multiple $3d$ configurations rather than a single integer valence \cite{mandal2018valence}; for Cr and Ti impurities in V$_2$O$_3$, dynamic correlations uncover an impurity-selective Mott transition driven by local symmetry breaking \cite{lechermann2018uncovering}; for oxygen vacancies in LaTiO$_3$, they preserve a robust Mott insulating state, contradicting the metallic electron-doping predicted by standard DFT\cite{souto2019dft+}; for the charge-transfer insulator NiO, dynamical many-body effects on the ligand $2p$ orbitals are essential to reproduce experimental spectral satellites\cite{lechermann2019interplay}.Together, these precedents establish that a many-body treatment is necessary to fully resolve the redox behavior and charge-transfer character of Fe-doped BaTiO$_3$ with $\mathrm{V_O}$.

In this study, the redox behavior of Fe impurities in BTO and its modulation by V$_{\mathrm{O}}$ formation are investigated via DFT-based many-body approaches. Two defect configurations are considered: the isolated dopant (Fe$_{\mathrm{Ti}}$) and the vacancy-compensated complex (Fe$_{\mathrm{Ti}}$V$_{\mathrm{O}}$), allowing intrinsic correlation effects to be disentangled from vacancy-induced contributions. The mixed-valence and spin-state character of the Fe $3d$ ground state are analyzed via atomic multiplet probabilities, while the redox response to V$_{\mathrm{O}}$ formation is quantified through orbital-resolved occupation matrices. Constrained DFT+U calculations over all symmetry-distinct $d^4$, $d^5$, and $d^6$ configurations are performed to establish a static mean-field reference. To further validate the predicted valence state, XPS and XAS are computed using a Wannier-based crystal-field and multiplet ligand field approach, demonstrating good agreement with recent experimental measurements. 

\section{Results and Discussion}\label{sec:results}

\subsection{Valence and spin fluctuations of Fe-doped BTO}\label{sec:fluctuations}

\begin{figure*}[!t] 
\centering
\includegraphics[width=\textwidth]{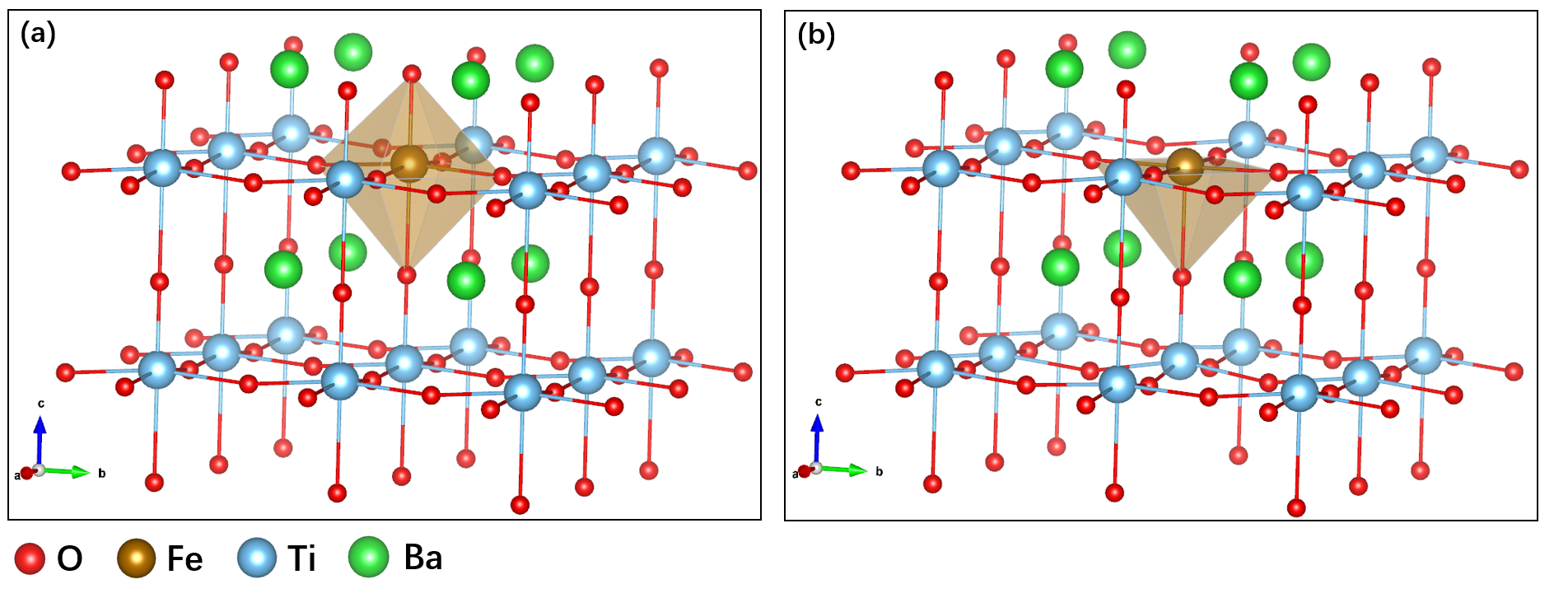}
\caption{Crystal structures of tetragonal Fe-substituted BTO supercells, optimized within DFT+U. (a) The isolated dopant Fe$_{\mathrm{Ti}}$ and (b) the $V_{\mathrm{O}}$-compensated complex Fe$_{\mathrm{Ti}}$V$_{\mathrm{O}}$. The Fe coordination polyhedron is highlighted in both panels: a full FeO$_6$ octahedron in (a) and a FeO$_5$ square pyramid in (b) following
removal of the apical oxygen. Both supercells retain the $P4mm$ symmetry along the $c$ axis.}
\label{fig:structures}
\end{figure*}

\begin{figure*}[!t]
\centering
\includegraphics[width=\textwidth]{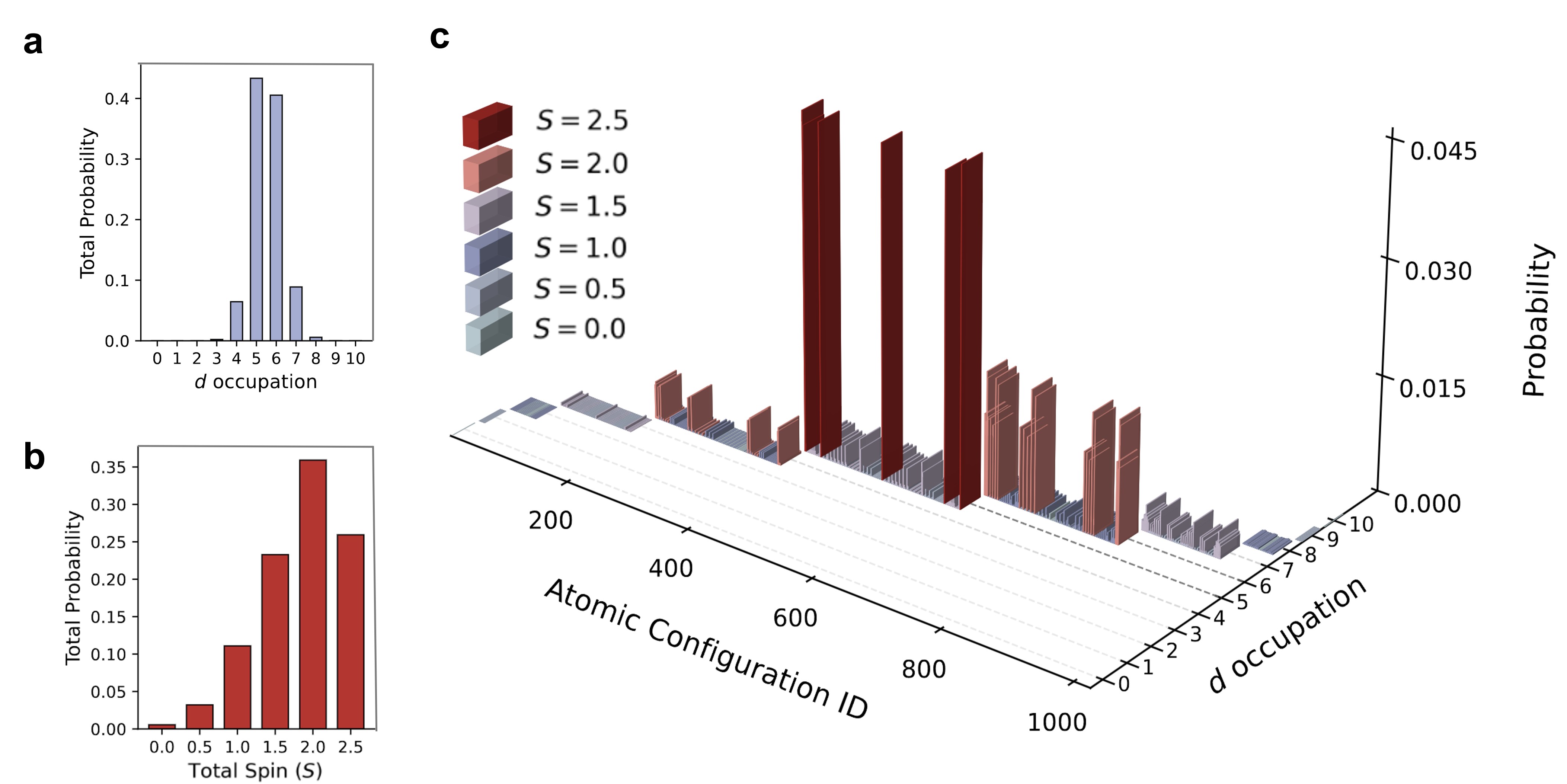}\\[4pt]
\includegraphics[width=\textwidth]{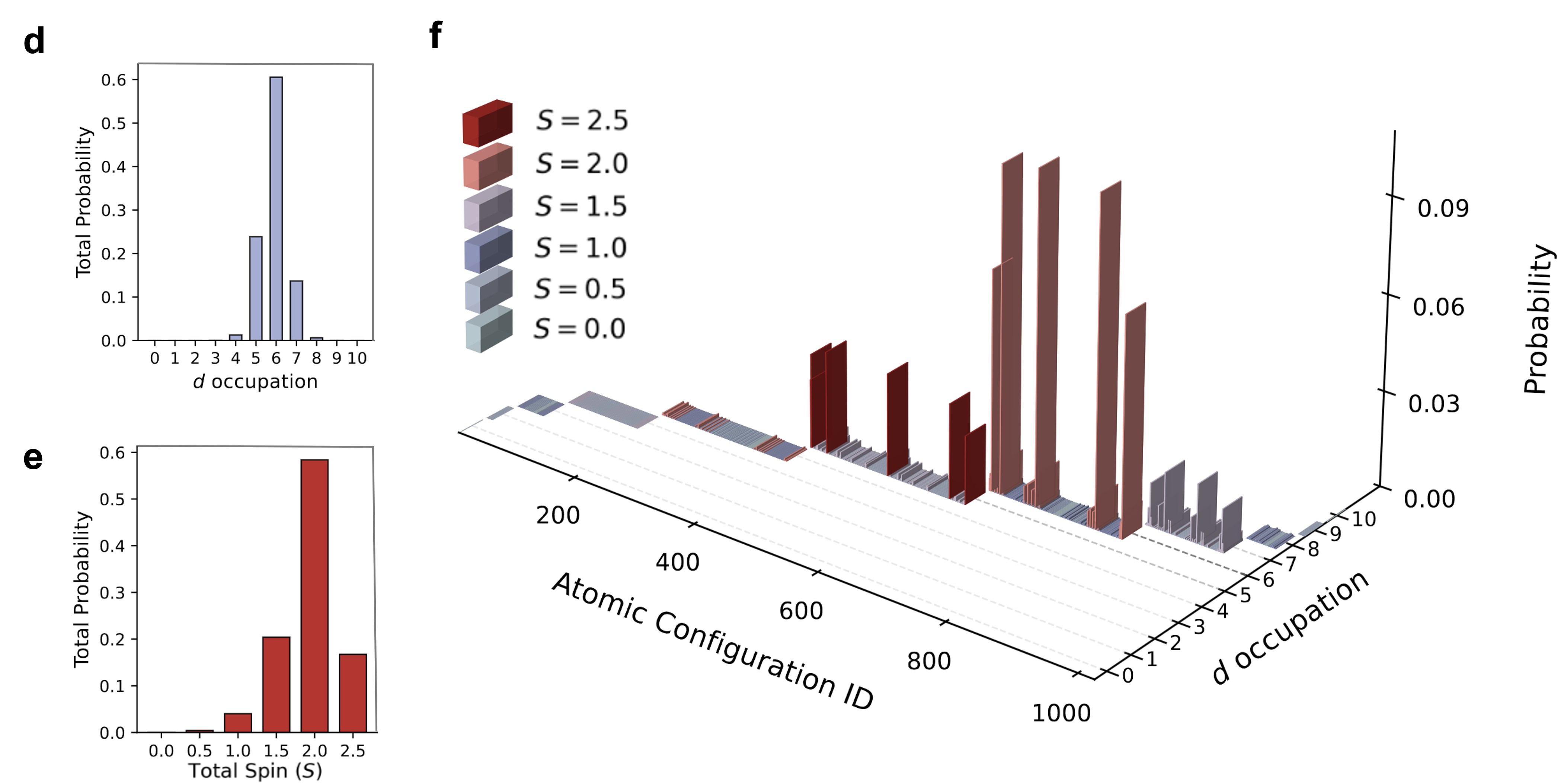}

\caption{DFT+DMFT computed atomic multiplet probabilities of the Fe $3d$ shell for Fe-substituted BTO treated in the paramagnetic state. The top row corresponds to the isolated dopant (Fe$_{\mathrm{Ti}}$) and the bottom row to the V$_{\mathrm{O}}$-compensated complex (Fe$_{\mathrm{Ti}}$V$_{\mathrm{O}}$). 
(a, d) Probability distribution of the total Fe $3d$ occupation $N$. 
(b, e) Probability distribution of the spin states $S$. 
(c, f) Probabilities of all $2^{10} = 1024$ individual atomic multiplets as a function of the atomic configuration index and occupation $N$, color-coded by the spin magnitude $S$.}
\label{fig:histograms}
\end{figure*}

The local valence state of the Fe impurity is first examined via the total Fe $3d$ orbital occupation $N_d$. For Fe$_{\mathrm{Ti}}$, $N_d \approx 5.54$ under the exact double-counting (DC) scheme~\cite{haule2015exact}, substantially larger than the $d^4$ configuration expected for a nominal Fe$^{4+}$ state, indicating a strong deviation from the ionic picture. Introducing compensating V$_{\mathrm{O}}$ drives a decisive further increase to $N_d \approx 5.9$--$6.0$ for Fe$_{\mathrm{Ti}}$V$_{\mathrm{O}}$, suggesting a pronounced reduction response to vacancy formation. The robustness of such results is confirmed by comparing different DC schemes and Coulomb interaction forms as detailed in Table S1: the choice between the density-density (Ising) and full rotationally Invariant Slater (Full) interactions have a negligible effect on $N_d$, while the fully localized limit (FLL)~\cite{czyzyk1994local} and nominal~\cite{haule2010dynamical,haule2014covalency} DC schemes yield slightly higher occupations than exact DC, a trend consistent with previous DFT+DMFT studies on Mn-doped BTO~\cite{mandal2018valence}. The overall physical picture of a strongly mixed-valence Fe$_{\mathrm{Ti}}$ that shifts decisively toward $d^6$ upon vacancy formation is therefore robust across all parameter choices tested.

The mixed valence nature of Fe impurities in BTO can be clearly obtained from atomic multiplet probabilities of various shells with different numbers of electrons obtained from DFT+DMFT calculations, as shown in Figs.~\ref{fig:histograms}(a,d). For Fe$_{\mathrm{Ti}}$, the distribution is broadly spread across the mixed-valence manifold: $\sim 7\%$ for $N=4$ (Fe$^{4+}$), $\sim 43\%$ for $N=5$ (Fe$^{3+}$), $\sim 40\%$ for $N=6$ (Fe$^{2+}$), and $\sim 10\%$ for $N=7$ (Fe$^{1+}$). Upon introducing the V$_{\mathrm{O}}$, the $N$-distribution for Fe$_{\mathrm{Ti}}$V$_{\mathrm{O}}$ shifts substantially: the weight at $N=4$ drops to a negligible $\sim 2\%$, while $N=6$ becomes the dominant configuration at $\sim 60\%$ and $N=5$ is reduced to $\sim 24\%$. This V$_{\mathrm{O}}$-driven redistribution is further evident in the comparison of the full multiplet landscapes (Fig.~\ref{fig:histograms}(c) and Fig.~\ref{fig:histograms}(f)): for Fe$_{\mathrm{Ti}}$, the $N=5$ and $N=6$ sectors each harbor scattered peaks of modest height ($<0.04$) with no single configuration exceeding $\sim 4\%$ probability, whereas for Fe$_{\mathrm{Ti}}$V$_{\mathrm{O}}$ the $N=6$ sector dominates with several high-probability peaks ($>0.1$) and $N=4$ is essentially absent. These findings demonstrate that the Fe $3d$ ground state cannot be described by a single integer valence, and that V$_{\mathrm{O}}$ formation effectively localises the valence toward $d^6$ (Fe$^{2+}$).

Turning now to the spin-state distribution $S$ as shown in Figs.~\ref{fig:histograms}(b,e). For Fe$_{\mathrm{Ti}}$, peaks at $S=2.5$ and $S=2.0$ are visible, yet the integrated probability for the fully polarised $S=2.5$ state is slightly below $30\%$, notably less than the total $N=5$ weight ($43\%$). This indicates that a significant fraction of Fe$^{3+}$ configurations do not adopt the high-spin state, resulting in an average spin projection $\langle S\rangle \approx 1.85$ and a local moment $\langle m \rangle \approx 3.7~\mu_B$, reduced from the theoretical maximum for a pure high-spin mixture. The multiplet landscape (Fig.~\ref{fig:histograms}(c)) reflects this: while the highest peaks carry $S =2.5$, a substantial background of intermediate and low-spin configurations ($S \leq 2.0$) is clearly present. For Fe$_{\mathrm{Ti}}$V$_{\mathrm{O}}$, the spin distribution sharpens markedly: $S =2$ becomes the defining feature of the spectrum (Fig.~\ref{fig:histograms}(e)), and the low-spin background is strongly suppressed in the multiplet landscape (Fig.~\ref{fig:histograms}(f)). The average moment remains similar ($\approx 3.8~\mu_B$), but the underlying distribution now points to a well-defined high-spin $d^6$ ($S=2$) character. Taken together, the valence and spin analyses indicate that V$_{\mathrm{O}}$ formation simultaneously localises the charge toward $d^6$ and narrows the spin fluctuation s toward $S =2$, consistent with a high-spin Fe$^{2+}$ ground state.

\begin{table*}[!t]
\centering
\caption{DFT+DMFT orbital occupation $N_d$ and average spin 
$\langle S\rangle$ for Fe$_\mathrm{Ti}$ and Fe$_\mathrm{Ti}$$V_\mathrm{O}$ 
under paramagnetic and ferromagnetic  frameworks. }
\label{tab:pm_fm}
\begin{tabular}{llcc}
\toprule
System & Magnetic Reference & $N_d$ & $\langle S\rangle$ \\
\midrule
\multirow{2}{*}{Fe$_\mathrm{Ti}$}
 & paramagnetic & 5.54 & 1.84 \\
 & ferromagnetic & 5.47 & 1.50 \\
\midrule
\multirow{2}{*}{Fe$_\mathrm{Ti}$$V_\mathrm{O}$}
 & paramagnetic & 5.88 & 1.93 \\
 & ferromagnetic & 5.67 & 1.86 \\
\bottomrule
\end{tabular}
\end{table*}

To further establish the robustness of the mixed-valence picture, ferromagnetic (FM) DFT+DMFT calculations are performed on the relaxed structures, though the experimentally observed magnetic moments in Fe-doped BTO are substantially reduced from the high-spin limit, typically well below 4 or 5 $\mu_B$ per Fe~\cite{xu2009room, ray2008high}. Table~\ref{tab:pm_fm} summarizes $N_d$ and $\langle S \rangle$ for both frameworks. Two observations are noteworthy. First, the mixed-valence character persists under FM ordering: $N_d$ remains non-integer in all cases, confirming that the ground state cannot be reduced to a single Fe valence regardless of the magnetic reference frame. Second, the introduction of $V_\mathrm{O}$ consistently increases $N_d$ in both frameworks, confirming the vacancy-driven charge localization toward $d^6$ identified earlier. The FM calculations yield somewhat lower $N_d$ and $\langle S \rangle$ than the paramagnetic ones, a difference consistent with the distinct treatment of spin fluctuations in the two frameworks. Crucially, the vacancy-induced enhancement of the local spin state is more pronounced in the FM framework: $\langle S \rangle$ increases from 1.50 to 1.86 upon $V_\mathrm{O}$ formation, a jump significantly larger than the corresponding paramagnetic shift (1.84 $\to$ 1.93), providing additional evidence that $V_\mathrm{O}$ can stabilize the high-spin $d^6$ configurations.

\subsection{Constrained DFT+U of Fe-doped BTO}\label{sec:conU}

\begin{figure}[htbp]
\centering
\includegraphics[width=\columnwidth]{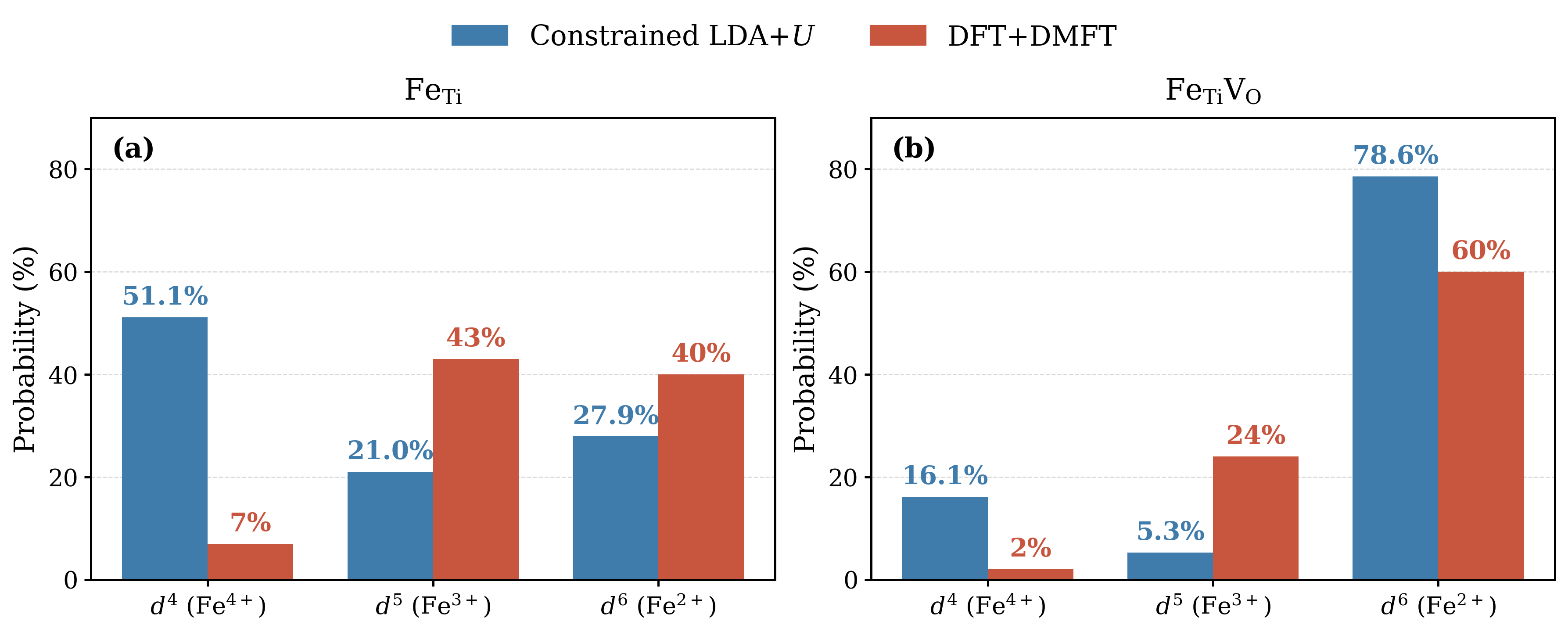}
\caption{Comparison of valence-state probabilities for Fe-substituted BTO
obtained from constrained DFT+U (blue) and DFT+DMFT (red), for the
uncompensated Fe$_{\mathrm{Ti}}$ (left) and vacancy-compensated
Fe$_{\mathrm{Ti}}$V$_{\mathrm{O}}$ (right) systems. Constrained DFT +U
probabilities are derived from Boltzmann distributions over 672 atomic
configurations at $T = 300$\,K; DFT+DMFT probabilities correspond to the
summed histogram weights in Fig.~\ref{fig:histograms}.}
\label{fig:conU_dmft}
\end{figure}

Such mixed-valence states can be intuitively understood based on constrained DFT+U calculations. Figure~\ref{fig:conU_dmft} compares the valence-state probability distributions obtained from constrained DFT+U (Boltzmann weights over 672 symmetry-distinct $d^4$, $d^5$, and $d^6$ configurations at $T = 300$\,K) against those from DFT+DMFT, for both Fe$_\mathrm{Ti}$ and Fe$_\mathrm{Ti}$$V_\mathrm{O}$. However, their quantitative distributions vary: constrained DFT+U assigns 51\%, 21\%, and 28\% to the $d^4$, $d^5$, and $d^6$ sectors, whereas DFT+DMFT spreads the weight to 7\%, 43\%, and 40\%, respectively. Within the static constrained DFT+U framework, the lowest-energy $d^4$ states are a pair of symmetry-related low-spin configurations, $\uparrow[d_{xy}, d_{yz}]\downarrow[d_{xy}, d_{xz}]$ and $\uparrow[d_{xy}, d_{xz}]\downarrow[d_{xy}, d_{yz}]$, capturing over 30\% of the total probability. Upon introducing a compensating V$_\mathrm{O}$, both methods predict a decisive shift toward the $d^6$ (Fe$^{2+}$) sector, which becomes dominant at ${\sim}79\%$ in constrained DFT+U and ${\sim}60\%$ in DFT+DMFT. The dominant static $d^6$ states are again two low-spin symmetry partners, $\uparrow[d_{xy}, d_{yz}, d_{3z^2-r^2}]\downarrow[d_{xy}, d_{xz}, d_{3z^2-r^2}]$ and $\uparrow[d_{xy}, d_{xz}, d_{3z^2-r^2}]\downarrow[d_{xy}, d_{yz}, d_{3z^2-r^2}]$, combining for over 50\% of the probability. While both methods capture the V$_\mathrm{O}$-driven shift toward Fe$^{2+}$, they yield distinct observations regarding the spin states and multiplet distributions. In both Fe$_\mathrm{Ti}$ and Fe$_\mathrm{Ti}$V$_\mathrm{O}$, the lowest-energy configurations in constrained DFT+U are predominantly low-spin. In contrast, the DFT+DMFT distributions are dominated by high-spin multiplets. Furthermore, for the uncompensated Fe$_\mathrm{Ti}$ system, constrained DFT+U retains a majority weight (51\%) in the nominal  Fe$^{4+}$($d^4$) state, whereas DFT+DMFT distributes the probability more evenly across the $d^5$ and $d^6$ states.  These differences highlight the necessity of a many-body treatment for a system with strong mixed-valence character. In addition, two inherent limitations of the constrained DFT+U procedure should be noted. First, the constrained energies include an artificial on-site penalty absent in the fully relaxed ground state, so the true energy ordering of configurations, particularly those with strong Fe–O covalent mixing, differs from the constrained values. Second, each configuration is described as a single Slater determinant with a fixed integer occupation, inherently neglecting the quantum superposition of multiplets that is central to the DMFT picture.

\subsection{Average fluctuating magnetic moment of Fe-doped BTO}
\label{sec:fluctuating_moment}

\begin{figure}[htbp]
\centering
\includegraphics[width=\columnwidth]{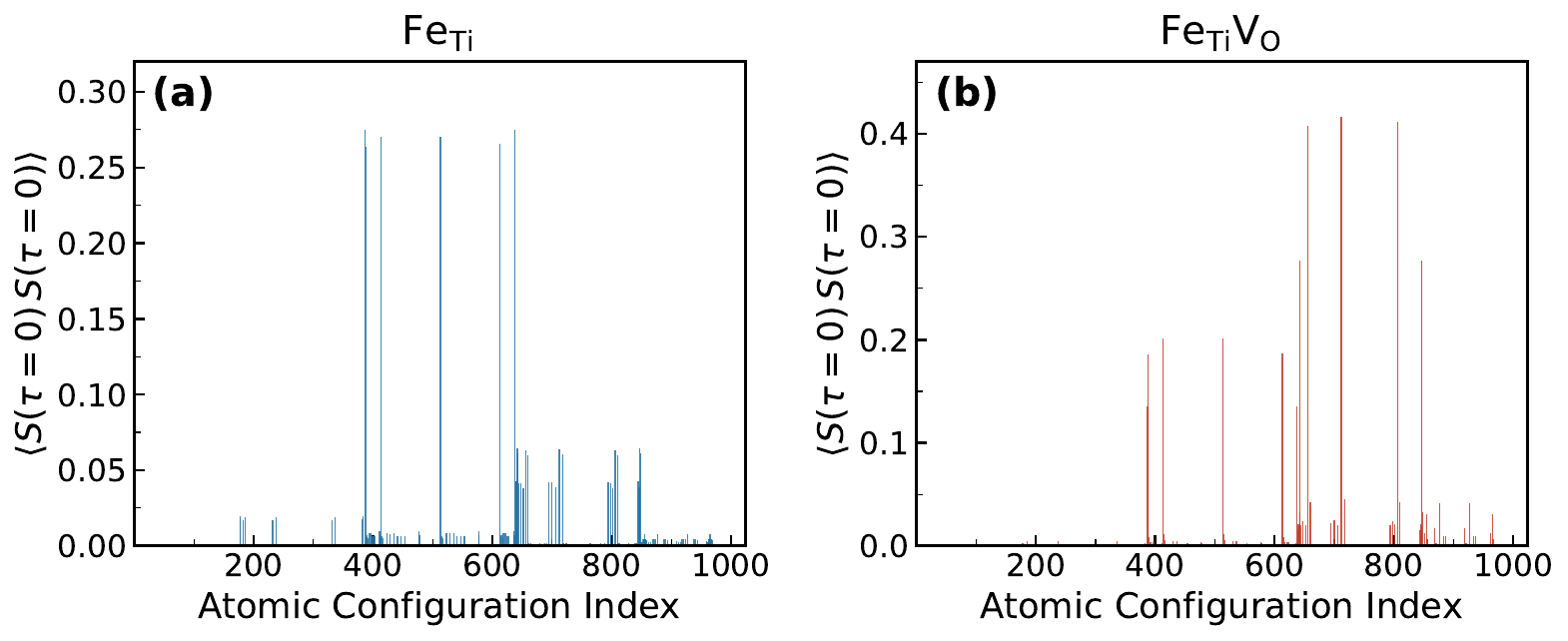}
\caption{DFT+DMFT computed squared fluctuating local moment
$\langle S(\tau{=}0)\,S(\tau{=}0)\rangle$ for Fe-substituted BTO
(a) without (Fe$_{\mathrm{Ti}}$) and (b) with compensating V$_{\mathrm{O}}$
(Fe$_{\mathrm{Ti}}$V$_{\mathrm{O}}$).}
\label{fig:squared_moment}
\end{figure}

The microscopic origin of the local magnetic moment can be further clarified by analyzing the time-averaged spin-spin correlation function, $\langle S(\tau=0)S(\tau=0) \rangle$, also referred to as the squared fluctuating local moment (Fig.~\ref{fig:squared_moment}). 
In the uncompensated Fe$_{\mathrm{Ti}}$ system (Fig.~\ref{fig:squared_moment}(a)), the distribution is broad. While prominent peaks with amplitudes $>0.3$ are observable (corresponding to the dominant high-spin $N=5$ multiplets identified in Fig.~\ref{fig:histograms}), they are accompanied by a dense landscape of low-amplitude peaks ($< 0.1$). This feature is indicative of the "noisy" magnetic ground state described in Sec. 3.1, where electrons hopping to ligands promote a diverse mixture of low-spin and fluctuating configurations. In contrast, for Fe$_{\mathrm{Ti}}$V$_{\mathrm{O}}$ (Fig.~\ref{fig:squared_moment}(b)), the magnetic spectral weight is notably cleaner. The number of peaks decreases markedly, and the background of low-amplitude fluctuations is suppressed. The distribution is dominated by high-amplitude peaks ($> 0.4$), concentrating the moment contributions on fewer, more probable high-spin configurations. This suggests that the V$_{\mathrm{O}}$ limits electron hopping channels, effectively filtering out the low-spin fluctuations and enhancing the magnetic coherence of the Fe center.

\begin{table*}[!t]
\centering
\caption{Temperature dependence of the squared fluctuating local moment 
($\langle S(\tau{=}0)S(\tau{=}0)\rangle$), orbital occupation ($N_d$), 
and average spin value ($\langle S \rangle$) for Fe-doped 
BTO calculated by DFT+DMFT.}
\label{tab:temp_dependence}
\begin{tabular}{lcccc}
\toprule
System & $T$ (K) & $\langle S(\tau{=}0)S(\tau{=}0)\rangle$ & $N_d$ & $\langle S \rangle$ \\ 
\midrule
\multirow{5}{*}{Fe$_{\mathrm{Ti}}$} 
 & 150  & 3.67 & 5.50 & 1.83 \\
 & 300  & 3.70 & 5.54 & 1.84 \\
 & 600  & 3.72 & 5.54 & 1.84 \\
 & 900  & 3.70 & 5.54 & 1.84 \\
 & 1200 & 3.71 & 5.54 & 1.84 \\
\midrule
\multirow{5}{*}{Fe$_{\mathrm{Ti}}$V$_{\mathrm{O}}$} 
 & 150  & 3.90 & 5.87 & 1.94 \\
 & 300  & 3.88 & 5.88 & 1.93 \\
 & 600  & 3.86 & 5.89 & 1.93 \\
 & 900  & 3.83 & 5.89 & 1.92 \\
 & 1200 & 3.81 & 5.90 & 1.91 \\
\bottomrule
\end{tabular}
\end{table*}

Finally, to assess whether these local electronic states persist under realistic processing conditions, we investigated the temperature dependence of the squared fluctuating local moment $\langle S(\tau{=}0)S(\tau{=}0)\rangle$, orbital occupation $N_d$, and $\langle S \rangle$ from 150 K to 1200 K (Table~\ref{tab:temp_dependence}). For the uncompensated Fe$_{\mathrm{Ti}}$ system, all three quantities remain remarkably stable across the entire 300--1200 K range, with $N_d \approx 5.54$ and $\langle S(\tau{=}0)S(\tau{=}0)\rangle \approx 3.70$ showing no systematic trend with temperature. The slight deviation observed at 150 K is attributed to improved numerical convergence of the CTQMC solver at lower temperatures, where double times of the Monte Carlo sweeps are set. For the Fe$_{\mathrm{Ti}}$V$_{\mathrm{O}}$ complex, a weak but systematic trend is observed: $\langle S(\tau{=}0)S(\tau{=}0)\rangle$ increases marginally with decreasing temperature while $N_d$ decreases slightly, consistent with the progressive suppression of high-occupation ($d^7$) thermal fluctuations as the system approaches its $d^6$ ground state. This overall thermal robustness indicates that the local moment is protected by an energy scale significantly larger than $k_B T$ even at 1200 K, consistent with the role of Hund's coupling J$_H$ in stabilizing local moments~\cite{werner2008spin,georges1996dynamical}. Such a robust magnetic moment provides a microscopic basis for experimental observations of high-temperature magnetic behavior in Fe-doped titanates~\cite{chakraborty2011defect,ganegoda2021role}.

\subsection{Orbital-Resolved Electronic Structure}\label{sec:orbital}

To figure out the origin of the changes of mixed valence nature for Fe impurities upon the introduction of V$_{\mathrm{O}}$, orbital occupations and the underlying crystal fields can be obtained via the DFT+U and DFT+DMFT electronic structure. In our paramagnetic DFT+DMFT caclulations, the occupation matrices in the basis $(d_{3z^2-r^2},\, d_{x^2-y^2},\, d_{xz},\, d_{yz},\, d_{xy})$ (under frozen-ion conditions) read:
\begin{align*}
\mathbf{n}(\mathrm{Fe_{Ti}}) &= (1.058,\ 1.047,\ 1.151,\ 1.151,\ 1.144), \\
\mathbf{n}(\mathrm{Fe_{Ti}V_O}) &= (1.459,\ 1.154,\ 1.054,\ 1.054,\ 1.077).
\end{align*}
That is, the total number of electrons is increases by $0.29\,e^-$, $i.e.$, from $N_d = 5.55$ to $5.84$, due to the charge transfer between Fe impurities and surrounding oxygen atoms. One noticeable orbital-resolved change is that the net charge on the $d_{3z^2-r^2}$ orbital is increased by $0.40$ electrons, whereas all $t_{2g}$ orbitals ($d_{xz}$, $d_{yz}$, $d_{xy}$) showing a modest decrease of $\sim\!0.07$--$0.10$ electrons. 
Structural relaxation ($i.e.$, optimization of atomic positions) causes only marginal modifications to the trend, where the corresponding occupation matrices read:
\begin{align*}
\mathbf{n}(\mathrm{Fe_{Ti}}) &= (1.058,\ 1.056,\ 1.141,\ 1.141,\ 1.135), \\
\mathbf{n}(\mathrm{Fe_{Ti}V_O}) &= (1.585,\ 1.140,\ 1.046,\ 1.051,\ 1.063).
\end{align*}
The $d_{3z^2-r^2}$ occupation increases by $0.53$ electrons, while the total number of electrons increases from  $N_d = 5.53$ to $N_d = 5.89$ by 0.35 electrons. The $t_{2g}$ occupations of the stoichiometric Fe$_{\mathrm{Ti}}$ cell are essentially unchanged for the frozen-ion and relaxed geometries.
Thus, it is suggested that the relocation of the $d_{3z^2-r^2}$ orbitals plays an essential role for the redox behavior of Fe impurities in BTO, which is consistent with the local symmetry breaking introduced by the apical oxygen removal, further analyzed in Sec.~\ref{sec:spectroscopy}.

\begin{figure}[htbp]
  \centering
  \includegraphics[width=\linewidth]{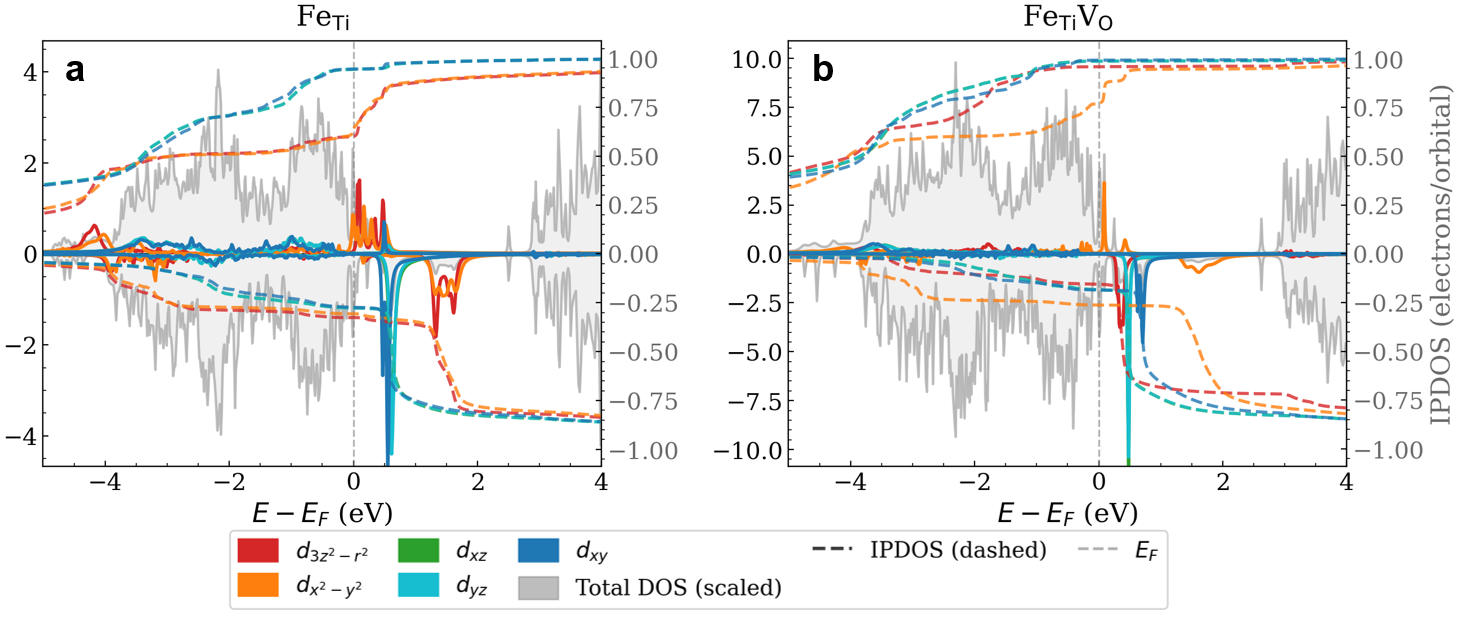}
  \caption{Spin-resolved Fe $3d$ spectral functions and integrated
    partial density of states (IPDOS) from FM DFT+DMFT for
    (a)~Fe$_{\mathrm{Ti}}$ and (b)~Fe$_{\mathrm{Ti}}$V$_{\mathrm{O}}$.
    Solid lines show the orbital-resolved PDOS
    $A_m(\omega) = -\frac{1}{\pi}\,\mathrm{Im}\,G_{mm}(\omega+i0^+)$
    with spin-up (spin-down) plotted positive (negative).
    Dashed lines show the corresponding cumulative IPDOS on the
    right axis. The grey shaded area is the total DOS (scaled).
    The vertical dashed line marks $E_F$.}
  \label{fig:pdos}
\end{figure}

As studied in our previous work~\cite{li2026origin}, the influence of magnetic ordering can be investigated via ferromagnetic (FM) DFT+DMFT calculations.
For the relaxed structures, the diagonal elements of the spin-resolved occupation matrices for Fe$_{\mathrm{Ti}}$ read:
\begin{align*}
\mathbf{n}^{\uparrow} &= (0.601,\ 0.628,\ 0.938,\ 0.938,\ 0.939), \\
\mathbf{n}^{\downarrow} &= (0.327,\ 0.308,\ 0.278,\ 0.278,\ 0.273),
\end{align*}
and for Fe$_{\mathrm{Ti}}$V$_{\mathrm{O}}$:
\begin{align*}
\mathbf{n}^{\uparrow} &= (0.955,\ 0.774,\ 0.984,\ 0.984,\ 0.989), \\
\mathbf{n}^{\downarrow} &= (0.157,\ 0.265,\ 0.185,\ 0.185,\ 0.188).
\end{align*}
For the isolated Fe$_{\mathrm{Ti}}$, the majority-spin $t_{2g}^{\uparrow}$ orbitals are nearly fully occupied ($\sim\!0.94$), while the $e_g^{\uparrow}$ orbitals are approximately half-filled ($\sim\!0.61$--$0.63$). This is consistent with the partial density of states (PDOS), as shown in Fig.~\ref{fig:pdos}(a), where unoccupied $d_{3z^2-r^2}$ and $d_{x^2-y^2}$ peaks in the majority spin channel are directly above the Fermi energy. The minority-spin orbitals are mostly unoccupied, $i.e.$, with partial occupations across all five orbitals around 0.3 electrons, resulting in $m \approx 2.58~\mu_B$. Upon introducing V$_{\mathrm{O}}$, the dominant changes take place in the $d_{3z^2-r^2}$ orbitals in both the majority-spin and minority-spin channels. On the one hand, the occupation of the $d_{3z^2-r^2}^{\uparrow}$ orbitals increases from $0.60$ to $0.95$ by $0.35$ electrons. On the other hand, although the occupation in the $d_{3z^2-r^2}^{\downarrow}$ orbital changes from $0.33$ to $0.16$ electrons, its energy is reduced, becoming comparable with that of the $t_{2g}$ orbitals in the minority-spin channel, as shown in Fig.~\ref{fig:pdos}(b). The magnitude of the local moments increases to $m \approx 3.71~\mu_B$, consistent with the V$_{\mathrm{O}}$-driven stabilization of the high-spin state observed in Sec.~\ref{sec:fluctuations}. The same orbital selectivity can be observed in the paramagnetic DFT+DMFT spectral functions (Fig.~S2), where, despite the absence of exchange splitting causing the in-gap features less distinct than in Fig.~\ref{fig:pdos}, the spectral weight of the $d_{3z^2-r^2}$ orbital below $E_F$ is markedly enhanced upon introducing V$_\mathrm{O}$. Therefore, we are convinced that the changes in the mixed valence nature and the redox behavior of Fe impurities can be attributed to the crystal field effects for the $d_{3z^2-r^2}$ orbitals driven by the appearance of $\text{V}_\text{O}$. 

The redox behavior and mixed valence nature of Fe impurites were also investigated in detail in Ref.~\cite{noguchi2020successive}, where a smaller U$_{\mathrm{Fe}\, d}$ = 2~eV together with an additional U$_{\mathrm{Ti}\, d}$ = 8~eV to open the host bandgap were used. This is different than what we used in our DFT+U and DMFT setting (U$_{\mathrm{Fe}\, d}$ = 5~eV, J$_H$ = 0.86~eV, no U on Ti d orbitals). To assess the consistency of our methodology with this reference, we performed DFT+U benchmark calculations using both parameter sets. Adopting the parameters used in our DMFT yields a V$_{\mathrm{O}}$-induced reorganization involving both $d_{xy}$ and $d_{3z^2-r^2}$, whereas adopting the parameters of Ref.~\cite{noguchi2020successive} reproduces the $d_{3z^2-r^2}$-only reorganization as reported there. Under the latter parameter setting, our DFT+U result agrees with Ref.~\cite{noguchi2020successive} on the dominant features: down-channel $d_{3z^2-r^2}$ selectivity and $d_{x^2-y^2}$ as the highest up-channel state upon V$_{\mathrm{O}}$ formation. Detailed comparisons are provided in the Fig.~S3. We now turn to the DMFT spectral functions, which extend this picture beyond the single-determinant DFT+U description. Figure~\ref{fig:pdos} shows the spin-resolved Fe $3d$ spectral functions from FM DFT+DMFT. For the isolated Fe$_{\mathrm{Ti}}$ (Fig.~\ref{fig:pdos}(a)), the down-channel $t_{2g}$ spectral weight extends close to $E_F$, in contrast to the placement of $\sim$1.8~eV above VBM in the pure Fe$^{3+}$ picture of Ref.~\cite{noguchi2020successive}. This downward shift is consistent with the substantial Fe$^{2+}$ weight ($\sim$40\%) identified in the multiplet histogram (Sec.~\ref{sec:fluctuations}). For Fe$_{\mathrm{Ti}}$V$_{\mathrm{O}}$ (Fig.~\ref{fig:pdos}(b)), all five $d$ orbitals appear as in-gap features. The Fe$^{3+}$--V$_{\mathrm{O}}$ picture of Ref.~\cite{noguchi2020successive} places $\{d_{xz}, d_{yz}, d_{3z^2-r^2}, d_{xy}\}$ as in-gap states with $d_{x^2-y^2}$ near the CBM, while the Fe$^{2+}$--V$_{\mathrm{O}}$ picture identifies $d_{3z^2-r^2}^{\downarrow}$ and $d_{x^2-y^2}^{\uparrow}$ as the dominant in-gap states. The full set of $d$-orbital in-gap features in our DMFT is naturally understood as the superposition of these two single-valence pictures, weighted by their multiplet probabilities. This unified mixed-valence picture emerges from a single charge-self-consistent DMFT calculation, without manually constraining the Fe valence to either Fe$^{2+}$ or Fe$^{3+}$.
\subsection{Spectroscopy Calculations}\label{sec:spectroscopy}
\begin{figure}[H]
    \centering
    \includegraphics[width=1\linewidth]{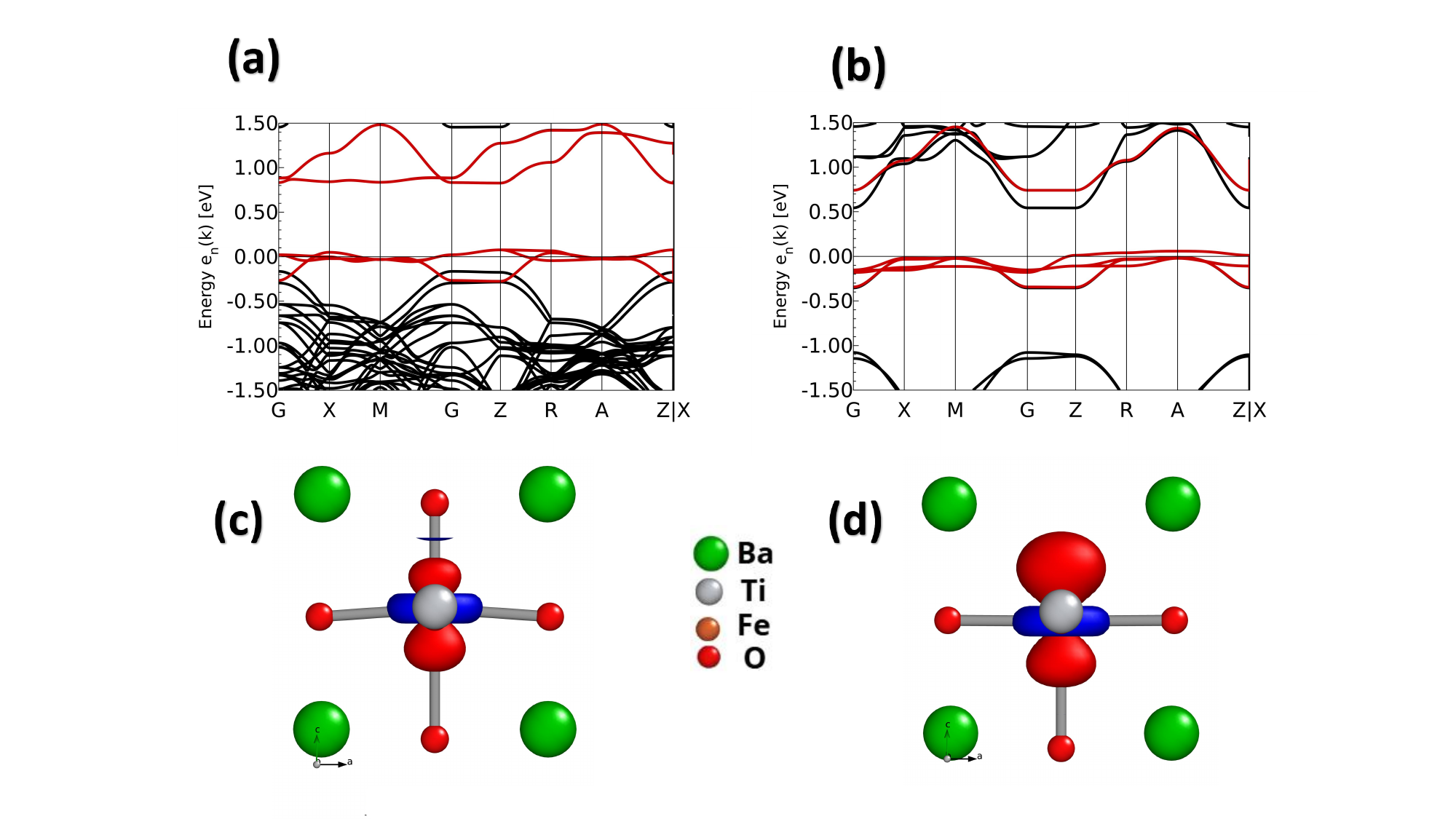}
\caption{
Wannier functions for Fe defects. 
(a) Wannier-interpolated bands (red) and DFT bands (black), projected onto the Fe $3d$ orbitals, for $\mathrm{Fe_{Ti}}$. 
(b) Same as in (a) for $\mathrm{Fe_{Ti}V_O}$ complex, with the Fermi energy set to zero. 
(c) Wannier function in the real space associated with the Fe $d_{z^2}$ state of $\mathrm{Fe_{Ti}}$. 
(d) Same as in (c) for $\mathrm{Fe_{Ti}V_O}$.
}
\label{fig:wannier}
\end{figure}

Turning now to the spectroscopic features of Fe impurities, the band structure and derived-Wannier functions for Fe-$3d$ orbitals are shown in Fig.~\ref{fig:wannier}. Obviously, for Fe$_{\mathrm{Ti}}$, the Fe-$3d$ bands dominate the states around the Fermi level, with partially occupied $t_{2g}$ bands and unoccupied $e_g$ bands (Fig.~\ref{fig:wannier}(a)). Upon the introduction of $\text{V}_\text{O}$, the $d_{3z^2-r^2}$ orbital is shifted downwards, leaving only the $d_{x^2-y^2}$ orbital unoccupied (Fig.~\ref{fig:wannier}(b)). It is noted that the Wannier functions are obtained for nonmagnetic calculations in the DFT (non-DFT+U) regime. Interestingly, such changes in the $k$-space can be visualized clearly via the real-space representation of the Wannier functions, as highlighted for the Fe $d_{3z^2-r^2}$orbital in Fig.~\ref{fig:wannier}(c,d). Note that the Wannier functions of the other Fe-$3d$ orbitals have only marginal changes, as shown in Fig.~S4.
 For Fe${\mathrm{Ti}}$ (Fig.~\ref{fig:wannier}(c)), the asymmetric shape of the Wannier function arises from local atomic displacements. Upon introducing $\mathrm{V_O}$ (Fig.~\ref{fig:wannier}(d)), the corresponding Wannier function exhibits an enhanced weight toward the $\mathrm{V_O}$ sites. Such a redistribution reflects overall reduced Fe--O hybridization, as evidenced by the hybridization function in Fig.~S6, together with increased localization of the $d_{3z^2-r^2}$ state, thereby favoring its enhanced occupation in agreement with the crystal-field analysis and DFT+DMFT results discussed above. 

 Correspondingly, the crystal fields of Fe-$3d$ orbitals are modified, which can be quantified by deriving the corresponding tight-binding Hamiltonians, obtained based on the Wannier construction using cutoff radii of 4.34 and 4.54~Bohr for Fe$_{\mathrm{Ti}}$
and $\mathrm{Fe_{Ti}V_O}$, respectively. 
The resulting on-site crystal-field
Hamiltonian for the Fe $3d$ basis ($d_{3z^2-r^2}$, $d_{x^2-y^2}$, $d_{xz}$, $d_{yz}$, $d_{xy}$) takes the form
\begin{equation}
\begin{aligned}
\mathrm{H}_\mathrm{Fe_{Ti}V_O}^{\mathrm{(3d)}} &=
\begin{pmatrix}
-0.045 & 0 & 0 & 0 & 0 \\
0 & 1.098 & 0 & 0 & 0 \\
0 & 0 & -0.081 & 0 & 0 \\
0 & 0 & 0 & -0.081 & 0 \\
0 & 0 & 0 & 0 & -0.103
\end{pmatrix}, \\[1em]
\mathrm{H}_\mathrm{Fe_{Ti}}^{(3d)} &=
\begin{pmatrix}
1.110 & 0 & 0 & 0 & 0 \\
0 & 1.168 & 0 & 0 & 0 \\
0 & 0 & -0.005 & 0 & 0 \\
0 & 0 & 0 & -0.005 & 0 \\
0 & 0 & 0 & 0 & -0.039
\end{pmatrix}.
\end{aligned}
\end{equation}
Obviously, the crystal fields of $3d$ orbitals for Fe$_{\mathrm{Ti}}$ show typical a $t_{2g}$-$e_g$ splitting of about 1.11 eV, with slight distortions caused by the tetragonal crystalline environments. A strong modification of the crystal-field splitting is obvious upon the introduction of the $\mathrm{V}_{\mathrm{O}}$, most notably a pronounced lowering of the $d_{\mathrm{3z^2-r^2}}$ orbital energy by 1.16 eV, from 1.110 eV to $-$0.045 eV. This is further verified by the DFT+DMFT charge density redistribution (Fig.~S5), which reveals obvious electron accumulation along the $z$-axis, slight accumulation in the $d_{x^2-y^2}$ channel, and tiny depletion in the $t_{2g}$ channels.

\begin{figure}[h]
    \centering
    
    \begin{subfigure}{0.45\textwidth}
        \centering
        \includegraphics[width=\linewidth]{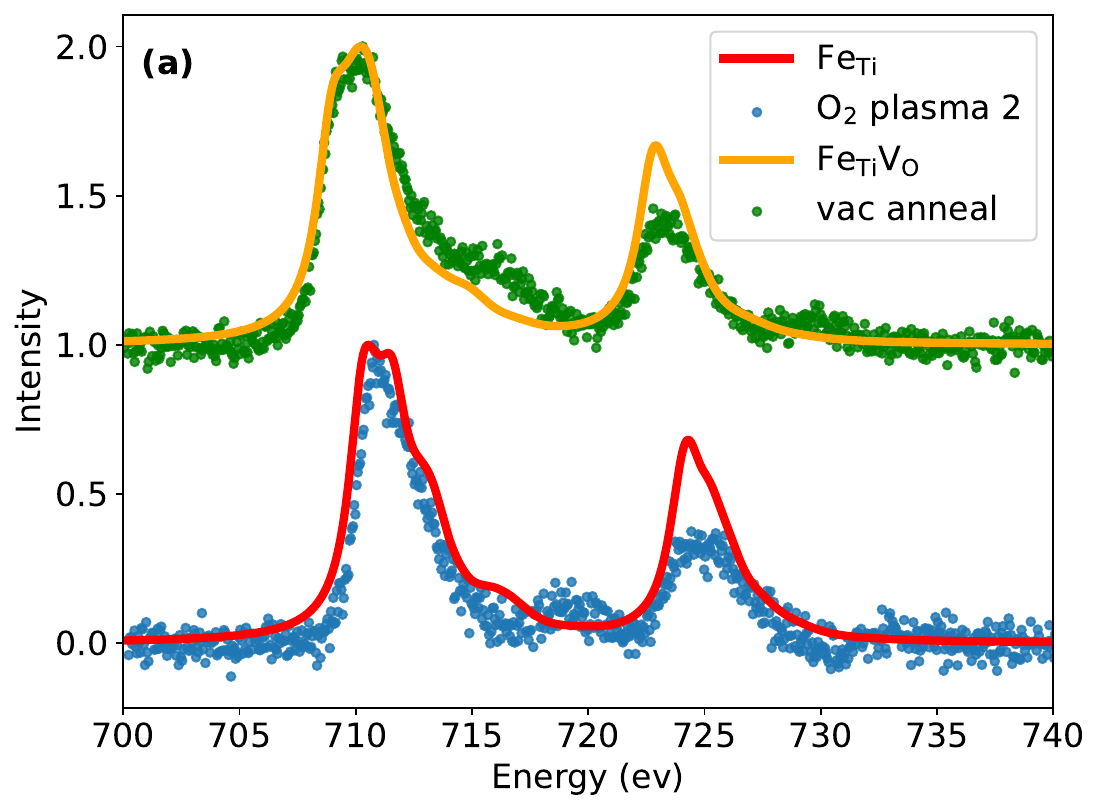}
     
    \end{subfigure}
    \begin{subfigure}{0.45\textwidth}
        \centering
        \includegraphics[width=\linewidth]{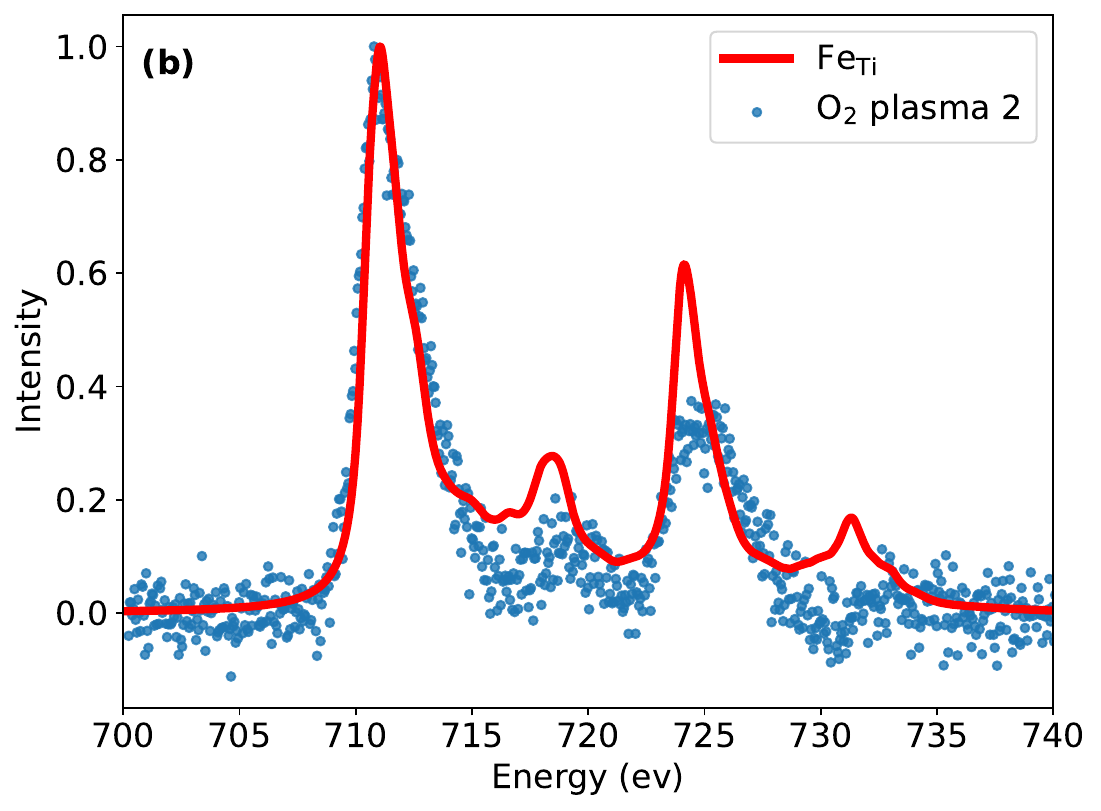}

    \end{subfigure}

    \caption{(a) X-ray photoelectron absortbion (XPS) compared with experimental data.
    The spectra are normalized to the maximum intensity.
    (b) X-ray photoemission spectra including charge-transfer effects for the $\mathrm{FeO_6}$ cluster.
    }
    \label{fig:spectraxps}
\end{figure}

Figure~\ref{fig:spectraxps}(a) shows the calculated Fe~2$p$ XPS spectra 
for Fe$_{\text{Ti}}$ and Fe$_{\text{Ti}}$V$_{\text{O}}$, compared against 
the experimental spectra of Ref.~\cite{chaoudhary2026direct}, where reducing 
(Ar/H$_{2}$) and oxidizing (O$_{2}$ plasma) conditions were used to 
selectively stabilize the Fe$^{2+}$ and Fe$^{3+}$ oxidation states. 
The calculated spectra reproduce the characteristic spectral fingerprints 
of both oxidation states with good agreement. The Fe$_{\text{Ti}}$ 
configuration captures the Fe$^{3+}$ spectral lineshape, while 
Fe$_{\text{Ti}}$V$_{\text{O}}$ reproduces the spectral weight redistribution 
and peak shift toward lower binding energies associated with Fe$^{2+}$, 
arising from reduced Fe--O hybridization and local crystal-field symmetry 
breaking. This correspondence provides direct microscopic support for oxygen 
vacancies as the driving mechanism behind Fe valence switching in BaTiO$_3$, 
and contrasts with negative charge-transfer oxides such as 
La$_{1-x}$Sr$_x$FeO$_3$, where oxygen vacancies have little impact on 
the cation valence and the XPS lineshape remains largely 
unchanged~\cite{wang2024recommended,mueller2015redox}.

To further assess the role of ligand states, we perform charge-transfer multiplet calculations for the FeO$_6$ cluster (Fig.~\ref{fig:spectraxps}(b)). Including ligand orbitals produces a pronounced satellite in the $\sim 717$--$720$~eV binding-energy range, consistent with the charge-transfer picture established for transition-metal oxides~\cite{xie2026redox}. The satellite arises from ground-state mixing between $d^{n}$ and $d^{n+1}\underline{L}$ configurations, where $\underline{L}$ denotes a ligand hole and the ligand weight controls the satellite intensity. Related Fe-based intercalation electrodes have recently been shown to undergo a crossover from positive- to negative-charge-transfer character, with ligand-centered holes strongly modifying the core-level spectra~\cite{ramachandran2026formal}. Our results therefore show that ligand charge-transfer effects extend beyond a purely crystal-field description, making the Fe $2p$ XPS satellite a sensitive marker of metal--ligand covalency.  We further perform XAS calculations Fig.~S8  using the derived crystal-field parameters, which predict distinct spectral changes that can be tested experimentally.

\section{Conclusion}\label{sec:conclusion}

We perform charge-self-consistent DFT+DMFT calculations on Fe-substituted BTO with and without $\text{V}_\text{O}$, as well as crystal field multiplet calculations on the core-level spectroscopies. The ground state of Fe impurities is found to be of strongly mixed-valence nature, with probabilities of $\sim$7\%, 43\%, and 40\% for the $d^4$, $d^5$, and $d^6$ configurations, respectively, with $N_d \approx 5.54$ and $\langle m \rangle \approx 3.7\,\mu_B$. Introduction of a compensating V$_{\mathrm{O}}$ drives the valence toward $d^6$ ($\sim$60\%), with $S =2$ being the dominant spin state($\sim$58\%). Detailed analysis on the orbital-resolved occupation matrices and PDOS reveal that V$_{\mathrm{O}}$ formation induces an orbital-selective charge redistribution dominated by that in the $d_{3z^2-r^2}$ orbital, with gains of $+0.40\,e^-$ and $+0.53\,e^-$ under both frozen-ion and relaxed conditions, respectively. This is valid for both paramagnetic and ferromagnetic DFT+DMFT calculations. Additionally, the local magnetic moment rises from 2.58 to 3.71\,$\mu_B$, consistent with the V$_{\mathrm{O}}$-driven stabilization of a high-spin $d^6$ state. The DFT+DMFT spectral functions indicate that all five $d$ orbitals in the minority-spin channel contribute as in-gap states for Fe$_{\mathrm{Ti}}$V$_{\mathrm{O}}$, with $d_{3z^2-r^2}^{\downarrow}$ showing the dominant orbital response upon V$_{\mathrm{O}}$ introduction. Furthermore, Wannier-derived crystal-field Hamiltonians with multiplet simulations are performed to evaluate the core-level spectroscopies. The Wannier functions for Fe$_{\mathrm{Ti}}$ and Fe$_{\mathrm{Ti}}V_O$ clearly confirm that $\text{V}_\text{O}$ induces a strong lowering of the crystal field for the $d_{3z^2-r^2}$ orbital and a redistribution of spectral weight in XPS/XAS, consistent with recent experimental measurements.
\section{Computational Details}\label{sec:methods}

We modeled Fe dopants in a $2 \times 2 \times 2$ supercell of the tetragonal phase of BTO at room temperature, with (Fe$_{\mathrm{Ti}}$-V$_{\mathrm{O}}$) and without (Fe$_{\mathrm{Ti}}$) an apical shortest-neighboring compensating V$_{\mathrm{O}}$. The position of V$_{\mathrm{O}}$ is selected based on the total energy, which is also consistent with the previous report\cite{noguchi2020successive}.  This corresponds to a 12.5\% Fe concentration on the Ti site. It is noted that such a high concentration should experimentally induce a phase transition into the hexagonal phase under annealed conditions, as demonstrated in~\cite{li2026origin}; we focus on tetragonal BTO, which hosts the ferroelectric functionality of interest\cite{cohen1992origin, scott2007applications}, the redox behavior of Fe impurities in hexagonal BTO will be saved for future study. The crystal structures were optimized using DFT+U as implemented in VASP~\cite{kresse1996efficiency}, utilizing the Perdew-Burke-Ernzerhof (PBE) generalized gradient approximation~\cite{perdew1996generalized}. A Coulomb interaction U = 5.0\,eV and Hund's coupling J$_H$ = 0.86\,eV were applied to the Fe $3d$ orbitals. Following structural optimization, we performed extensive constrained DFT+U calculations using the occupation matrix control method~\cite{allen2014occupation}. Unlike standard DFT+U which typically converges to a single ground state, we systematically explored the energy landscape of specific valence configurations. We enumerated the possible states corresponding to nominal Fe$^{4+}$ ($3d^4$), Fe$^{3+}$ ($3d^5$), and Fe$^{2+}$ ($3d^6$) configurations. This involved calculating the total energies for all symmetry-distinct permutations of $d$-electron occupancies within the cubic harmonic basis, resulting in 672 considered configurations (the sum of combinatorial possibilities $C_{10}^4$, $C_{10}^5$, and $C_{10}^6$). The probability $P_i$ of each configuration $i$ was then derived using the Boltzmann distribution $P_i \propto \exp(-E_i/k_B T)$ at $T=300$ K, providing a static mean-field comparison among different valences.

To accurately capture the dynamic quantum fluctuations, we subsequently employed DFT+DMFT methods~\cite{kotliar2006electronic, haule2010dynamical}, considering both the paramagnetic and ferromagnetic states of the Fe dopants. The many-body calculations were performed using the WIEN2k code~\cite{blaha2020wien2k} and its eDMFT implementation~\cite{haule2010dynamical}, utilizing the continuous-time quantum Monte Carlo (CTQMC) method~\cite{haule2007quantum} as the impurity solver. To ensure consistency, the same interaction parameters were used: U = 5.0\,eV, consistent with charge-self-consistent DFT+DMFT calculations for Fe-based systems including iron pnictides and iron selenides~\cite{yin2011magnetism, liu2018robust}, and J$_H$ = 0.86\,eV adopted from DFT+DMFT studies of Fe in oxygen-coordinated environments~\cite{greenberg2018pressure}, appropriate for the octahedral oxide coordination of the Fe$_{\mathrm{Ti}}$ 
substitutional site. The resulting effective interaction U$_{\mathrm{eff}}$ = U - J$_H$ = 4.14\,eV  lies within the typical range established for Fe $3d$ states across a variety of host environments~\cite{meng2016density,grau2006electronic}. We also note that previous studies on Mn impurities in BTO have demonstrated that the orbital occupation remains robust against variations of U within a reasonable range~\cite{mandal2018valence}. We used a Monkhorst-Pack $k$-point mesh of at least $4\times4\times4$ and a total of $40$ million Monte Carlo steps per iteration to ensure high precision. The sensitivity of our results was tested against different double counting (DC) schemes, specifically the fully localized limit (FLL)~\cite{czyzyk1994local} and the ``exact'' double counting scheme~\cite{haule2015exact}. Finally, the orbital-resolved spectral functions, $A_m(\omega) = -\frac{1}{\pi}\,\mathrm{Im}\,G_{mm}(\omega + i0^+)$, were evaluated using the real-frequency axis self-energy obtained via the maximum entropy method~\cite{gubernatis1991quantum}.

To evaluate the core-level spectroscopies, we applied the crystal/ligand field multiplet approach as implemented in the quanty code \cite{haverkort2012multiplet}. The noninteracting atomic Hamiltonian was obtained based on the Wannier parameterization of DFT electronic structure computed using the FPLO code \cite{koepernik2023symmetry}.
The Wannier projection windows were optimized to achieve an accurate representation of the DFT band structure, {\it i.e.}, $d_{xy}$ in $[-0.11,\,0.059]$~eV, $d_{yz}$ and $d_{xz}$ in $[-0.11,\,0.12]$~eV, $d_{3z^2-r^2}$ in $[0.12,\,0.75]$~eV, and $d_{x^2-y^2}$ in $[0.75,\,1.42]$~eV for Fe$_{\mathrm{Ti}}$. Whereas for Fe$_{\mathrm{Ti}}V_{\mathrm{O}}$, modified energy windows were employed, {\it i.e.}, the $d_{xy}$, $d_{yz}$, $d_{xz}$, and $d_{3z^2-r^2}$ orbitals were projected within $[-0.4,\,0.15]$~eV, while the $d_{x^2-y^2}$ orbital was treated separately in the range $[0.7,\,1.06]$~eV (Fig S7). Furthermore, to capture the Fe-$3d$ and O-$2p$ hybridization effects for Fe$_{\mathrm{Ti}}$, an additional projection including both Fe-$3d$ and O-$2p$ orbitals was performed within an extended energy window of $[-5.5,\,1.534]$~eV. The XPS and XAS spectra were calculated from the imaginary part of the Green's function
$G(\omega) = \langle \Psi | T^+ (z - H)^{-1} T | \Psi \rangle$,
where $(z - H)^{-1}$ denotes the resolvent of the atomic Hamiltonian, $T$ is the transition
operator coupling core and valence states, and $z = \omega + E_0 + i\Gamma/2$, with
$E_0 = \langle \Psi | H | \Psi \rangle$ the ground-state energy and $\Gamma$ the intrinsic
lifetime broadening. The multiplet atomic Hamiltonian, constructed within the DFT+multiplet
approach using the Quanty code~\cite{haverkort2012multiplet}, was defined as
$H = H^{\mathrm{DFT}} - {}^{m}H_{U}^{dd} + \sum_{a=d,p} \left( H_{U}^{da} + H_{l.s.}^{a}
+ H_{\mathrm{ex}} \right)$
where $H^{\mathrm{DFT}}$ was the single-particle Hamiltonian obtained from DFT via a
$p$--$d$ Wannier basis~\cite{koepernik2023symmetry}, and the sum runs over the $3d$ shell
($a=d$) and the $2p$ core shell ($a=p$). To avoid double counting, the mean-field
Hartree--Fock-like part of the $3d$--$3d$ Coulomb interaction, ${}^{m}H_{U}^{dd}$, was
subtracted following the Slater-integral parametrization~\cite{haverkort2012multiplet}, and
the full Coulomb interaction $H_{U}^{da}$ was reintroduced together with spin--orbit coupling
$H_{l.s.}^{a}$ and the exchange field $H_{\mathrm{ex}}$. To facilitate comparison with
experiment, the spectra were aligned in binding energy, normalized to unity, and convoluted
with a Gaussian of $1.4\,\mathrm{eV}$ for XPS and $2.4\,\mathrm{eV}$ for XAS to account
for instrumental resolution and lifetime effects.

\backmatter
 
\section*{Declarations}
 
\noindent\textbf{Supplementary information:} 
\noindent Supplementary information accompanies this paper.
 
\noindent\textbf{Acknowledgements:} 
\noindent This work was supported by the Deutsche Forschungsgemeinschaft,
Project-ID 463184206 -- SFB 1548. We gratefully acknowledge computing time on the
high-performance computer Lichtenberg at NHR4CES at TU Darmstadt (project p0026451)
and at RWTH Aachen University (project p0024024), funded by the Federal Ministry of
Education and Research and the state governments participating on the basis of the
GWK resolutions (\url{https://www.nhr-verein.de/unsere-partner}).
 
\noindent\textbf{Competing interests:} 
\noindent The authors declare no competing interests.
 
\noindent\textbf{Data and code availability:} 
\noindent All data needed to reproduce the findings are
available upon reasonable request from the corresponding author.
 
\noindent\textbf{Author contributions:}
\noindent Z.L.\ and H.Ze.\ contributed equally.
Z.L.: software, investigation, visualization, formal analysis, writing -- original draft.
H.Ze.: software, investigation, visualization, formal analysis, writing -- original draft.
H.W.: methodology, supervision.
R.X.: methodology, supervision.
H.B.Z.: conceptualization, formal analysis, supervision, funding acquisition, writing -- review and editing.

\bibliography{ref}

@article{kotliar2006electronic,
  title={Electronic structure calculations with dynamical mean-field theory},
  author={Kotliar, Gabriel and Savrasov, Sergej Y and Haule, Kristjan and Oudovenko, Viktor S and Parcollet, O and Marianetti, CA},
  journal={Reviews of Modern Physics},
  volume={78},
  number={3},
  pages={865--951},
  year={2006},
  publisher={APS}
}

@article{chikada2010analysis,
  title={Analysis of local environment of Fe ions in hexagonal BaTiO3},
  author={Chikada, Shunsuke and Hirose, Kazuyuki and Yamamoto, Tomoyuki},
  journal={Japanese journal of applied physics},
  volume={49},
  number={9R},
  pages={091502},
  year={2010}
}

@article{langhammer2020incorporation,
  title={On the incorporation of iron into hexagonal barium titanate: II. Magnetic moment, electron paramagnetic resonance (EPR) and optical transmission},
  author={Langhammer, HT and Walther, T and B{\"o}ttcher, R and Ebbinghaus, SG},
  journal={Journal of Physics: Condensed Matter},
  volume={32},
  number={38},
  pages={385702},
  year={2020},
  publisher={IOP Publishing}
}

@article{bottcher2018incorporation,
  title={On the incorporation of iron into hexagonal barium titanate: I. electron paramagnetic resonance (EPR) study},
  author={B{\"o}ttcher, R and Langhammer, HT and K{\"u}cker, S and Eisenschmidt, C and Ebbinghaus, SG},
  journal={Journal of Physics: Condensed Matter},
  volume={30},
  number={42},
  pages={425701},
  year={2018},
  publisher={IOP Publishing}
}

@article{padchasri2021crystal,
  title={Crystal structure and XANES study of Fe-substituted Barium Titanate ceramics prepared by conventional solid-state technique},
  author={Padchasri, J and Triamnak, N and Sareein, T and Jutimoosik, J and Tongsaeng, S and Bootchanont, A and Kidkhunthod, P and Rujirawat, S and Manyum, P and Yimnirun, R},
  journal={Radiation Physics and Chemistry},
  volume={188},
  pages={109657},
  year={2021},
  publisher={Elsevier}
}

@article{chaoudhary2026direct,
  title={Direct determination of the Fe3+/2+ charge-transition level in BaTiO3 and isovalently substituted Ba0. 82Ca0. 18Ti0. 92Zr0. 08O3 by x-ray photoelectron spectroscopy},
  author={Chaoudhary, Savita and Paulik, Anna M and Bertelmann, Niklas and Sudarikov, Denis and Hu, Pengcheng and Lohaus, Katharina NS and Gossel, Lisanne and Larsson, Melissa A and Ali, Hebatallah and Blume, Raoul and others},
  journal={Journal of Applied Physics},
  volume={139},
  number={9},
  year={2026},
  publisher={AIP Publishing}
}

@article{lechermann2019interplay,
  title={Interplay of charge-transfer and Mott-Hubbard physics approached by an efficient combination of self-interaction correction and dynamical mean-field theory},
  author={Lechermann, Frank and K{\"o}rner, Wolfgang and Urban, Daniel F and Els{\"a}sser, Christian},
  journal={Physical Review B},
  volume={100},
  number={11},
  pages={115125},
  year={2019},
  publisher={APS}
}

@article{souto2019dft+,
  title={DFT+ DMFT study of oxygen vacancies in a Mott insulator},
  author={Souto-Casares, Jaime and Spaldin, Nicola A and Ederer, Claude},
  journal={Physical Review B},
  volume={100},
  number={8},
  pages={085146},
  year={2019},
  publisher={APS}
}

@article{lechermann2018uncovering,
  title={Uncovering the mechanism of the impurity-selective Mott transition in paramagnetic V 2 O 3},
  author={Lechermann, Frank and Bernstein, Noam and Mazin, II and Valent{\'\i}, Roser},
  journal={Physical review letters},
  volume={121},
  number={10},
  pages={106401},
  year={2018},
  publisher={APS}
}

@article{haule2007quantum,
  title={Quantum Monte Carlo impurity solver for cluster dynamical mean-field theory and electronic structure calculations with adjustable cluster base},
  author={Haule, Kristjan},
  journal={Physical Review B—Condensed Matter and Materials Physics},
  volume={75},
  number={15},
  pages={155113},
  year={2007},
  publisher={APS}
}

@article{haule2010dynamical,
  title={Dynamical mean-field theory within the full-potential methods: Electronic structure of CeIrIn 5, CeCoIn 5, and CeRhIn 5},
  author={Haule, Kristjan and Yee, Chuck-Hou and Kim, Kyoo},
  journal={Physical Review B—Condensed Matter and Materials Physics},
  volume={81},
  number={19},
  pages={195107},
  year={2010},
  publisher={APS}
}

@article{werner2008spin,
  title={Spin freezing transition and non-Fermi-liquid self-energy in a three-orbital model},
  author={Werner, Philipp and Gull, Emanuel and Troyer, Matthias and Millis, Andrew J},
  journal={Physical Review Letters},
  volume={101},
  number={16},
  pages={166405},
  year={2008},
  publisher={APS}
}

@article{maso2006electrical,
  title={Electrical properties of Fe-doped BaTiO 3},
  author={Maso, N and Beltran, H and Cordoncillo, E and Escribano, P and West, AR},
  journal={Journal of Materials Chemistry},
  volume={16},
  number={17},
  pages={1626--1633},
  year={2006},
  publisher={Royal Society of Chemistry}
}

@article{nguyen2011tetragonal,
  title={Tetragonal and hexagonal polymorphs of BaTi1- xFexO3- $\delta$ multiferroics using x-ray and Raman analyses},
  author={Nguyen, Ha M and Dang, NV and Chuang, Pei-Yu and Thanh, TD and Hu, Chih-Wei and Chen, Tsan-Yao and Lam, VD and Lee, Chih-Hao and Hong, LV},
  journal={Applied Physics Letters},
  volume={99},
  number={20},
  year={2011},
  publisher={AIP Publishing}
}

@article{perdew1996generalized,
  title={Generalized gradient approximation made simple},
  author={Perdew, John P and Burke, Kieron and Ernzerhof, Matthias},
  journal={Physical review letters},
  volume={77},
  number={18},
  pages={3865},
  year={1996},
  publisher={APS}
}

@article{mandal2018valence,
  title={Valence and spin fluctuations in the Mn-doped ferroelectric BaTiO 3},
  author={Mandal, Subhasish and Cohen, RE and Haule, K},
  journal={Physical Review B},
  volume={98},
  number={7},
  pages={075155},
  year={2018},
  publisher={APS}
}

@article{greenberg2018pressure,
  title={Pressure-induced site-selective Mott insulator-metal transition in Fe 2 O 3},
  author={Greenberg, Eran and Leonov, Ivan and Layek, Samar and Konopkova, Zuzana and Pasternak, Moshe P and Dubrovinsky, Leonid and Jeanloz, Raymond and Abrikosov, Igor A and Rozenberg, Gregory Kh},
  journal={Physical Review X},
  volume={8},
  number={3},
  pages={031059},
  year={2018},
  publisher={APS}
}

@article{meng2016density,
  title={When density functional approximations meet iron oxides},
  author={Meng, Yu and Liu, Xing-Wu and Huo, Chun-Fang and Guo, Wen-Ping and Cao, Dong-Bo and Peng, Qing and Dearden, Albert and Gonze, Xavier and Yang, Yong and Wang, Jianguo and others},
  journal={Journal of chemical theory and computation},
  volume={12},
  number={10},
  pages={5132--5144},
  year={2016},
  publisher={ACS Publications}
}

@article{grau2006electronic,
  title={Electronic structure and magnetic coupling in Fe Sb O 4: A DFT study using hybrid functionals and GGA+ U methods},
  author={Grau-Crespo, Ricardo and Cor{\`a}, Furio and Sokol, Alexey A and de Leeuw, Nora H and Catlow, C Richard A},
  journal={Physical Review B—Condensed Matter and Materials Physics},
  volume={73},
  number={3},
  pages={035116},
  year={2006},
  publisher={APS}
}

@article{czyzyk1994local,
  title={Local-density functional and on-site correlations: The electronic structure of La 2 CuO 4 and LaCuO 3},
  author={Czy{\.z}yk, MT and Sawatzky, GA},
  journal={Physical Review B},
  volume={49},
  number={20},
  pages={14211},
  year={1994},
  publisher={APS}
}

@article{haule2015exact,
  title={Exact double counting in combining the dynamical mean field theory and the density functional theory},
  author={Haule, Kristjan},
  journal={Physical review letters},
  volume={115},
  number={19},
  pages={196403},
  year={2015},
  publisher={APS}
}

@article{allen2014occupation,
  title={Occupation matrix control of d-and f-electron localisations using DFT+ U},
  author={Allen, Jeremy P and Watson, Graeme W},
  journal={Physical Chemistry Chemical Physics},
  volume={16},
  number={39},
  pages={21016--21031},
  year={2014},
  publisher={Royal Society of Chemistry}
}

@article{kresse1996efficiency,
  title={Efficiency of ab-initio total energy calculations for metals and semiconductors using a plane-wave basis set},
  author={Kresse, Georg and Furthm{\"u}ller, J{\"u}rgen},
  journal={Computational materials science},
  volume={6},
  number={1},
  pages={15--50},
  year={1996},
  publisher={Elsevier}
}

@article{haule2014covalency,
  title={Covalency in transition-metal oxides within all-electron dynamical mean-field theory},
  author={Haule, Kristjan and Birol, Turan and Kotliar, Gabriel},
  journal={Physical Review B},
  volume={90},
  number={7},
  pages={075136},
  year={2014},
  publisher={APS}
}

@article{ganegoda2021role,
  title={Role of Fe Doping on Local Structure and Electrical and Magnetic Properties of PbTiO3},
  author={Ganegoda, Hasitha and Mukherjee, Soham and Ma, Beihai and Olive, Daniel T and McNeely, James H and Kaduk, James A and Terry, Jeff and Rensmo, Hakan and Segre, Carlo U},
  journal={The Journal of Physical Chemistry C},
  volume={125},
  number={22},
  pages={12342--12354},
  year={2021},
  publisher={ACS Publications}
}

@article{chakraborty2011defect,
  title={Defect-induced magnetism: Test of dilute magnetism in Fe-doped hexagonal BaTiO 3 single crystals},
  author={Chakraborty, Tanushree and Ray, Sugata and Itoh, Mitsuru},
  journal={Physical Review B—Condensed Matter and Materials Physics},
  volume={83},
  number={14},
  pages={144407},
  year={2011},
  publisher={APS}
}

@article{rani2016structural,
  title={Structural, electrical, magnetic and magnetoelectric properties of Fe doped BaTiO3 ceramics},
  author={Rani, Alka and Kolte, Jayant and Vadla, Samba Siva and Gopalan, Prakash},
  journal={Ceramics International},
  volume={42},
  number={7},
  pages={8010--8016},
  year={2016},
  publisher={Elsevier}
}

@article{upadhyay2011enhanced,
  title={Enhanced photoelectrochemical response of BaTiO3 with Fe doping: experiments and first-principles analysis},
  author={Upadhyay, Sumant and Shrivastava, Jaya and Solanki, Anjana and Choudhary, Surbhi and Sharma, Vidhika and Kumar, Pushpendra and Singh, Nirupama and Satsangi, Vibha R and Shrivastav, Rohit and Waghmare, Umesh V and others},
  journal={The Journal of Physical Chemistry C},
  volume={115},
  number={49},
  pages={24373--24380},
  year={2011},
  publisher={ACS Publications}
}

@article{blaha2020wien2k,
  title={WIEN2k: An APW+ lo program for calculating the properties of solids},
  author={Blaha, Peter and Schwarz, Karlheinz and Tran, Fabien and Laskowski, Robert and Madsen, Georg KH and Marks, Laurence D},
  journal={The Journal of chemical physics},
  volume={152},
  number={7},
  year={2020},
  publisher={AIP Publishing}
}

@article{gubernatis1991quantum,
  title={Quantum Monte Carlo simulations and maximum entropy: Dynamics from imaginary-time data},
  author={Gubernatis, James E and Jarrell, Mark and Silver, RN and Sivia, DS},
  journal={Physical Review B},
  volume={44},
  number={12},
  pages={6011},
  year={1991},
  publisher={APS}
}

@article{cohen1992origin,
  title={Origin of ferroelectricity in perovskite oxides},
  author={Cohen, Ronald E},
  journal={Nature},
  volume={358},
  number={6382},
  pages={136--138},
  year={1992},
  publisher={Nature Publishing Group UK London}
}

@article{scott2007applications,
  title={Applications of modern ferroelectrics},
  author={Scott, JF},
  journal={science},
  volume={315},
  number={5814},
  pages={954--959},
  year={2007},
  publisher={American Association for the Advancement of Science}
}

@article{dietl2010ten,
  title={A ten-year perspective on dilute magnetic semiconductors and oxides},
  author={Dietl, Tomasz},
  journal={Nature materials},
  volume={9},
  number={12},
  pages={965--974},
  year={2010},
  publisher={Nature Publishing Group UK London}
}

@article{spaldin2005renaissance,
  title={The renaissance of magnetoelectric multiferroics},
  author={Spaldin, Nicola A and Fiebig, Manfred},
  journal={Science},
  volume={309},
  number={5733},
  pages={391--392},
  year={2005},
  publisher={American Association for the Advancement of Science}
}

@article{liu2018robust,
  title={Robust and clean Majorana zero mode in the vortex core of high-temperature superconductor (Li 0.84 Fe 0.16) OHFeSe},
  author={Liu, Qin and Chen, Chen and Zhang, Tong and Peng, Rui and Yan, Ya-Jun and Wen, Chen-Hao-Ping and Lou, Xia and Huang, Yu-Long and Tian, Jin-Peng and Dong, Xiao-Li and others},
  journal={Physical Review X},
  volume={8},
  number={4},
  pages={041056},
  year={2018},
  publisher={APS}
}

@article{yin2011magnetism,
  title={Magnetism and charge dynamics in iron pnictides},
  author={Yin, ZP and Haule, K and Kotliar, G},
  journal={Nature physics},
  volume={7},
  number={4},
  pages={294--297},
  year={2011},
  publisher={Nature Publishing Group UK London}
}

@article{ray2008high,
  title={High temperature ferromagnetism in single crystalline dilute Fe-doped Ba Ti O 3},
  author={Ray, Sugata and Mahadevan, Priya and Mandal, Suman and Krishnakumar, SR and Kuroda, Carlos Seiti and Sasaki, T and Taniyama, Tomoyasu and Itoh, Mitsuru},
  journal={Physical Review B—Condensed Matter and Materials Physics},
  volume={77},
  number={10},
  pages={104416},
  year={2008},
  publisher={APS}
}

@article{noguchi2020successive,
  title={Successive redox-mediated visible-light ferrophotovoltaics},
  author={Noguchi, Yuji and Taniguchi, Yuki and Inoue, Ryotaro and Miyayama, Masaru},
  journal={Nature Communications},
  volume={11},
  number={1},
  pages={966},
  year={2020},
  publisher={Nature Publishing Group UK London}
}

@article{keith2004synthesis,
  title={Synthesis and characterisation of doped 6H-BaTiO3 ceramics},
  author={Keith, Gillian M and Rampling, Mandy J and Sarma, K and Alford, Neil Mc and Sinclair, DC},
  journal={Journal of the European Ceramic Society},
  volume={24},
  number={6},
  pages={1721--1724},
  year={2004},
  publisher={Elsevier}
}

@article{xu2009room,
  title={Room-temperature ferromagnetism and ferroelectricity in Fe-doped BaTiO 3},
  author={Xu, B and Yin, KB and Lin, J and Xia, YD and Wan, XG and Yin, J and Bai, XJ and Du, J and Liu, ZG},
  journal={Physical Review B—Condensed Matter and Materials Physics},
  volume={79},
  number={13},
  pages={134109},
  year={2009},
  publisher={APS}
}

@article{mangalam2009multiferroic,
  title={Multiferroic properties of nanocrystalline BaTiO3},
  author={Mangalam, RVK and Ray, Nirat and Waghmare, Umesh V and Sundaresan, A and Rao, CNR},
  journal={Solid State Communications},
  volume={149},
  number={1-2},
  pages={1--5},
  year={2009},
  publisher={Elsevier}
}

@article{adeagbo2019theoretical,
  title={Theoretical investigation of iron incorporation in hexagonal barium titanate},
  author={Adeagbo, Waheed A and Ben Hamed, Hichem and Nayak, Sanjeev K and B{\"o}ttcher, Rolf and Langhammer, Hans T and Hergert, Wolfram},
  journal={Physical Review B},
  volume={100},
  number={18},
  pages={184108},
  year={2019},
  publisher={APS}
}

@article{li2026origin,
  title   = {Origin of the tetragonal-to-hexagonal phase transitions 
             in {Fe}-doped {BaTiO$_3$}},
  author  = {Li, Zhiyuan and Xie, Ruiwen and Zhang, Hongbin},
  journal = {arXiv preprint arXiv:2603.22571},
  year    = {2026}
}

@article{nakayama2001theoretical,
  title={Theoretical prediction of magnetic properties of Ba (Ti1-xMx) O3 (M= Sc, V, Cr, Mn, Fe, Co, Ni, Cu)},
  author={Nakayama, Hiroyuki Nakayama Hiroyuki and Katayama-Yoshida, Hiroshi Katayama-Yoshida Hiroshi},
  journal={Japanese journal of applied physics},
  volume={40},
  number={12B},
  pages={L1355},
  year={2001},
  publisher={IOP Publishing}
}

@article{mikulska2015x,
  title={X-ray Absorption Spectroscopy Studies of the Room-Temperature Ferromagnetic Fe-Doped 6H--BaTiO3},
  author={Mikulska, Iuliia and Valant, Matjaz and Ar{\v{c}}on, Iztok and Lisjak, Darja},
  journal={Journal of the American Ceramic Society},
  volume={98},
  number={4},
  pages={1156--1161},
  year={2015},
  publisher={Wiley Online Library}
}

@article{nossa2015effects,
  title={Effects of manganese addition on the electronic structure of BaTiO 3},
  author={Nossa, JF and Naumov, Ivan I and Cohen, RE},
  journal={Physical Review B},
  volume={91},
  number={21},
  pages={214105},
  year={2015},
  publisher={APS}
}

@article{dang2011structural,
  title={Structural, optical and magnetic properties of polycrystalline BaTi1- xFexO3 ceramics},
  author={Dang, NV and Thanh, TD and Hong, LV and Lam, VD and others},
  journal={Journal of Applied Physics},
  volume={110},
  number={4},
  year={2011},
  publisher={AIP Publishing}
}

@article{jartych2018mossbauer,
  title={M{\"o}ssbauer spectroscopy studies of Fe-doped BaTiO3 ceramics},
  author={Jartych, El{\.z}bieta and Pikula, T and Garbarz-Glos, Barbara and Panek, R},
  journal={Acta Phys. Pol. A},
  volume={134},
  pages={1058--1062},
  year={2018}
}

@article{anderson1961localized,
  title={Localized magnetic states in metals},
  author={Anderson, Philip Warren},
  journal={Physical Review},
  volume={124},
  number={1},
  pages={41},
  year={1961},
  publisher={APS}
}

@article{hubbard1963electron,
  title={Electron correlations in narrow energy bands},
  author={Hubbard, John},
  journal={Proceedings of the Royal Society of London. Series A. Mathematical and Physical Sciences},
  volume={276},
  number={1365},
  pages={238--257},
  year={1963},
  publisher={The Royal Society London}
}

@article{georges1996dynamical,
  title={Dynamical mean-field theory of strongly correlated fermion systems and the limit of infinite dimensions},
  author={Georges, Antoine and Kotliar, Gabriel and Krauth, Werner and Rozenberg, Marcelo J},
  journal={Reviews of modern physics},
  volume={68},
  number={1},
  pages={13},
  year={1996},
  publisher={APS}
}

@article{haverkort2012multiplet,
  title={Multiplet ligand-field theory using Wannier orbitals},
  author={Haverkort, MW and Zwierzycki, M and Andersen, OK},
  journal={Physical Review B—Condensed Matter and Materials Physics},
  volume={85},
  number={16},
  pages={165113},
  year={2012},
  publisher={APS}
}

@article{wang2024recommended,
  title={Recommended strategies for quantifying oxygen vacancies with X-ray photoelectron spectroscopy},
  author={Wang, Jiayue and Mueller, David N and Crumlin, Ethan J},
  journal={Journal of the European Ceramic Society},
  volume={44},
  number={15},
  pages={116709},
  year={2024},
  publisher={Elsevier}
}

@article{koepernik2023symmetry,
  title={Symmetry-conserving maximally projected Wannier functions},
  author={Koepernik, K and Janson, O and Sun, Yan and Van Den Brink, J},
  journal={Physical Review B},
  volume={107},
  number={23},
  pages={235135},
  year={2023},
  publisher={APS}
}

@article{xie2026redox,
  title={Redox chemistry of LiCoO2, LiNiO2, and LiNi1/3Mn1/3Co1/3O2 cathodes: Deduced via XPS, DFT+ DMFT, and charge transfer multiplet simulations},
  author={Xie, Ruiwen and Mellin, Maximilian and Jaegermann, Wolfram and Hofmann, Jan P and de Groot, Frank MF and Zhang, Hongbin},
  journal={Nano Energy},
  pages={111747},
  year={2026},
  publisher={Elsevier}
}

@article{ramachandran2026formal,
  title={A formal FeIII/V redox couple in an intercalation electrode},
  author={Ramachandran, Hari and Mu, Edward W and Lomeli, Eder G and Braun, Augustin and Goto, Masato and Hsu, Kuan H and Liu, Jue and Jiang, Zhelong and Lim, Kipil and Busse, Grace M and others},
  journal={Nature Materials},
  volume={25},
  number={1},
  pages={91--99},
  year={2026},
  publisher={Nature Publishing Group UK London}
}

@article{mueller2015redox,
  title={Redox activity of surface oxygen anions in oxygen-deficient perovskite oxides during electrochemical reactions},
  author={Mueller, David N and Machala, Michael L and Bluhm, Hendrik and Chueh, William C},
  journal={Nature communications},
  volume={6},
  number={1},
  pages={6097},
  year={2015},
  publisher={Nature Publishing Group UK London}
}

\end{document}


\begin{center}
    {\LARGE \textbf{Supplementary Information}}\\[1.5em]
    {\large \textbf{Redox behaviour of Fe impurities in BaTiO$_3$ based on
    many-body calculations}}\\[1.5em]
    {\normalsize
        Zhiyuan Li$^{\dagger,1}$,\;
        Hamza Zerdoumi$^{\dagger,1}$,\;
        Hao Wang$^{1}$,\;
        Ruiwen Xie$^{1}$,\;
        Hongbin Zhang$^{1,*}$
    }\\[0.8em]
    {\small
        $^1$Institute of Materials Science, Technische Universität Darmstadt,
        64287 Darmstadt, Germany\\[0.4em]
        $^\dagger$These authors contributed equally to this work.\\
        $^*$Corresponding author.
        E-mail: \href{mailto:hongbin.zhang@tu-darmstadt.de}%
        {hongbin.zhang@tu-darmstadt.de}
    }
\end{center}

\vspace{2em}
\hrule
\vspace{2em}


\begin{table}[htbp]
\centering
\caption{DFT+DMFT computed occupation ($N_d$) of the Fe $3d$ orbital in
Fe-doped BaTiO$_3$ without (Fe$_{\mathrm{Ti}}$) and with
(Fe$_{\mathrm{Ti}}$V$_{\mathrm{O}}$) a compensating V$_{\mathrm{O}}$.
Results are compared across different double counting (DC) and Coulomb
interaction schemes~\cite{czyzyk1994local, haule2010dynamical,
haule2014covalency, haule2015exact}.}
\label{tab:occupation}
\begin{tabular}{lllc}
\toprule
System & DC Scheme & Coulomb & $N_d$ \\
\midrule
\multirow{5}{*}{Fe$_{\mathrm{Ti}}$}
 & Exact   & Full  & 5.54 \\
 & Exact   & Ising & 5.55 \\
 & FLL     & Full  & 5.71 \\
 & FLL     & Ising & 5.72 \\
 & Nominal & Ising & 5.71 \\
\midrule
\multirow{5}{*}{Fe$_{\mathrm{Ti}}$V$_{\mathrm{O}}$}
 & Exact   & Full  & 5.89 \\
 & Exact   & Ising & 5.89 \\
 & FLL     & Full  & 5.99 \\
 & FLL     & Ising & 6.00 \\
 & Nominal & Ising & 6.00 \\
\bottomrule
\end{tabular}
\end{table}


\begin{figure*}[htbp]
\centering
\includegraphics[width=\textwidth]{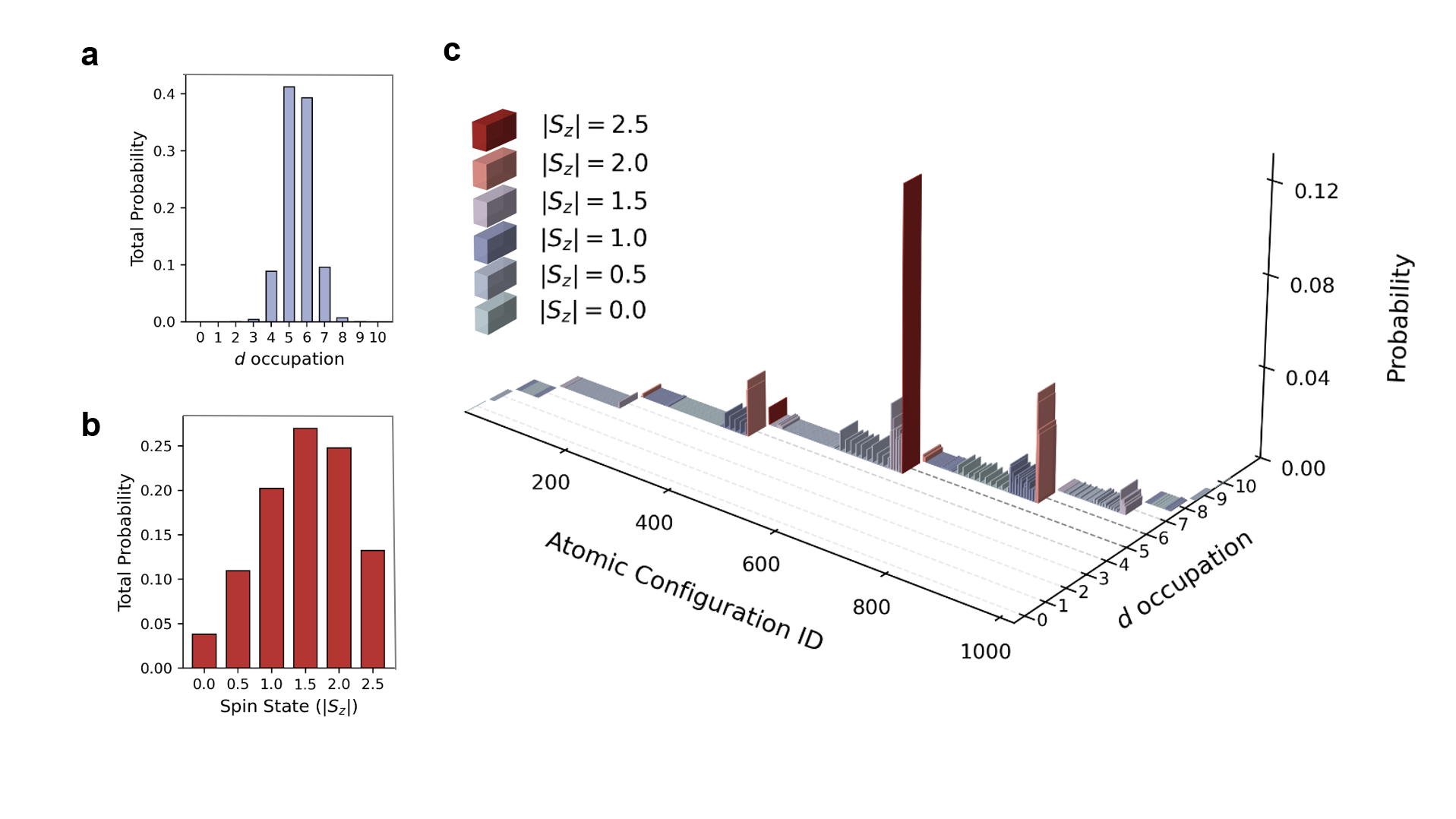}\\[4pt]
\includegraphics[width=\textwidth]{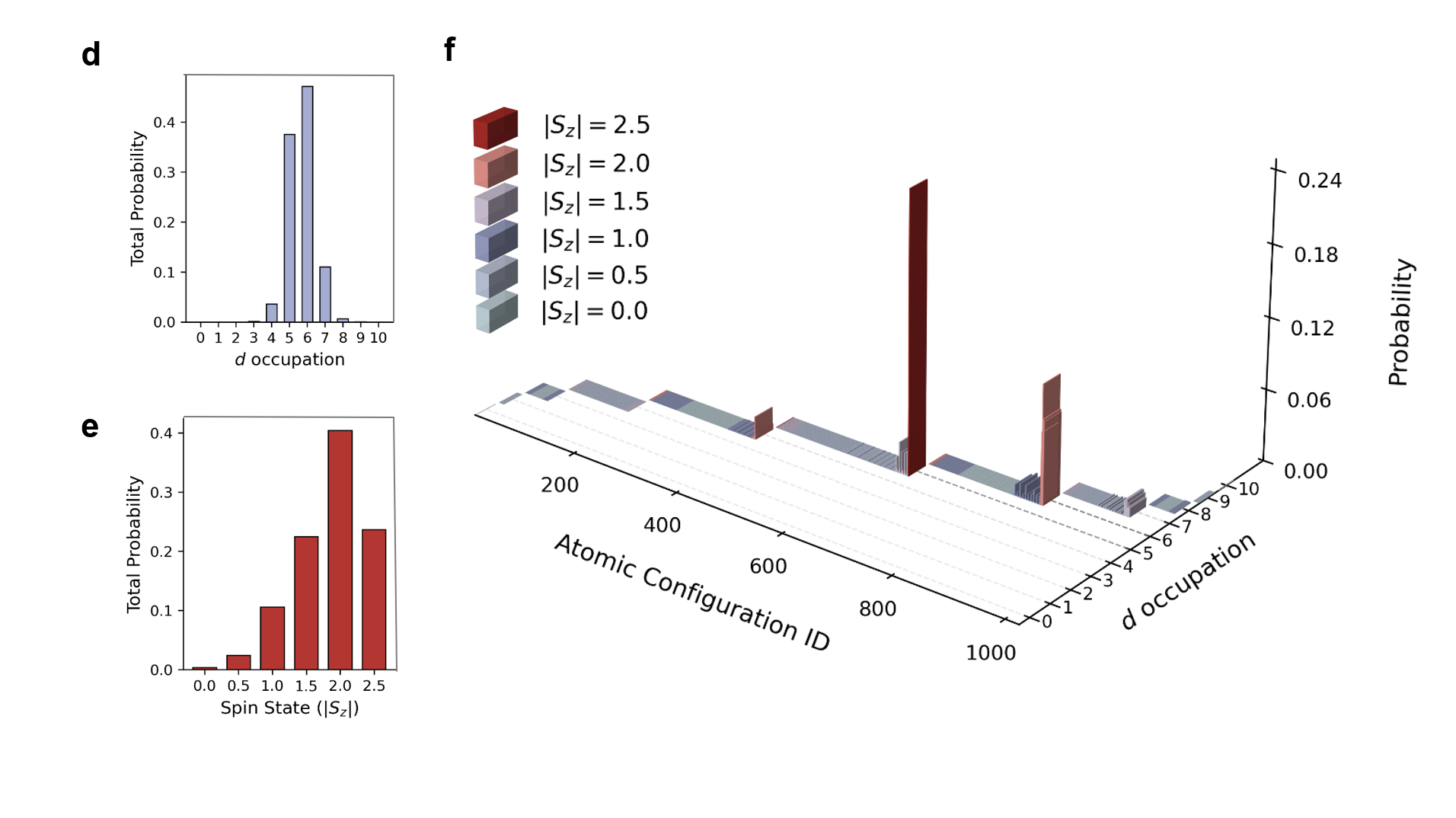}
\caption{DFT+DMFT computed atomic multiplet probabilities of the Fe $3d$ shell
for Fe-substituted BTO treated in the ferromagnetic state. The top row
corresponds to the isolated dopant (Fe$_{\mathrm{Ti}}$) and the bottom row to
the V$_{\mathrm{O}}$-compensated complex (Fe$_{\mathrm{Ti}}$V$_{\mathrm{O}}$).
(a, d) Probability distribution of the total Fe $3d$ occupation $N$.
(b, e) Probability distribution of the spin states $|S_z|$.
(c, f) Probabilities of all $2^{10} = 1024$ individual atomic multiplets as a
function of the atomic configuration index and occupation $N$, color-coded by
the spin states $|S_z|$.}
\label{fig:histograms_FM}
\end{figure*}


\begin{figure}[H]
  \centering
  \includegraphics[width=\linewidth]{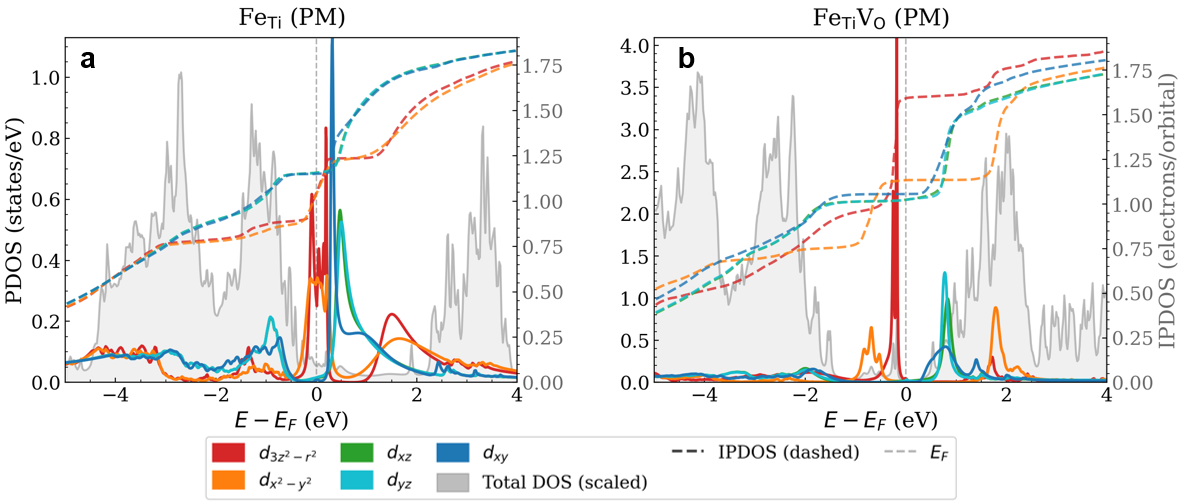}
  \caption{Fe $3d$ spectral functions and integrated partial density of states
    (IPDOS) from paramagnetic DFT+DMFT for (a)~Fe$_{\mathrm{Ti}}$ and
    (b)~Fe$_{\mathrm{Ti}}$V$_{\mathrm{O}}$.
    Solid lines show the orbital-resolved PDOS
    $A_m(\omega) = -\frac{1}{\pi}\,\mathrm{Im}\,G_{mm}(\omega+i0^+)$.
    Dashed lines show the corresponding cumulative IPDOS (spin-summed) on the
    right axis. The grey shaded area is the total DOS (scaled). The vertical
    dashed line marks $E_F$.}
  \label{fig:pdos_pm}
\end{figure}


\begin{figure}[H]
\centering
\includegraphics[width=\textwidth]{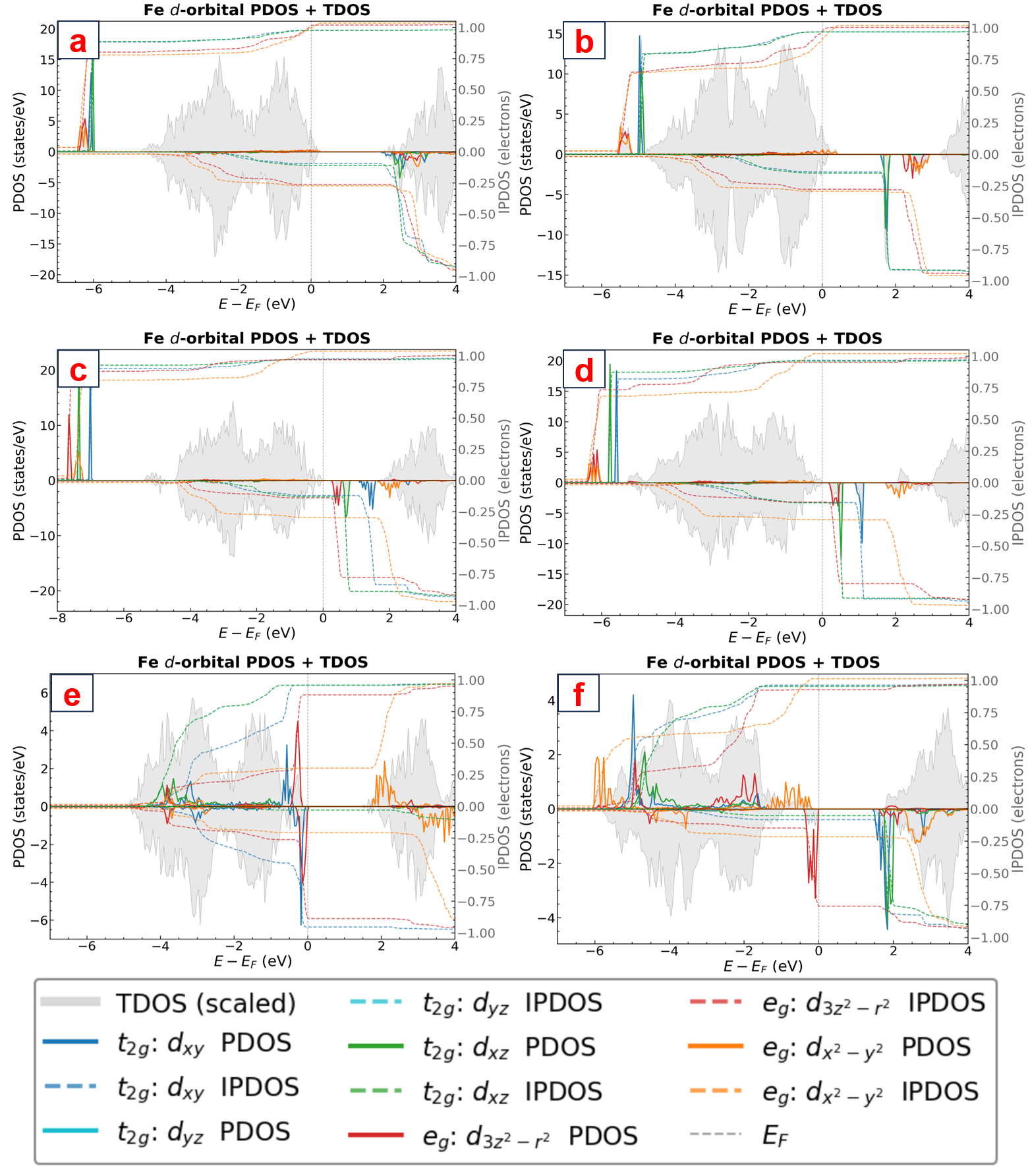}
\caption{Fe $3d$ orbital-resolved PDOS and IPDOS of Fe-substituted BTO under
two DFT$+U$ parameter sets. Rows correspond to (top) Fe$^{3+}$, (middle)
Fe$^{3+}$--V$_{\mathrm{O}}$, and (bottom) Fe$^{2+}$--V$_{\mathrm{O}}$.
(a, c, e) DFT$+U$ calculated using our DMFT parameters
($U_{\mathrm{Fe}\,d} = 5$~eV, $J_H = 0.86$~eV, no $U$ on Ti).
(b, d, f) DFT$+U$ calculated using the parameters of
Ref.~\cite{noguchi2020successive}
($U_{\mathrm{Fe}\,d} = 2$~eV, $U_{\mathrm{Ti}\,d} = 8$~eV).
In all panels, solid lines show the orbital-resolved PDOS and dashed lines the
corresponding cumulative IPDOS (right axis); spin-up (spin-down) channels are
plotted positive (negative). The grey shaded area is the total DOS (scaled).
The vertical dashed line marks $E_F$. For direct comparison with the reference
PDOS, see Fig.~4b, d, f of Ref.~\cite{noguchi2020successive}.}
\label{fig:Ucomparison}
\end{figure}


\begin{figure}[H]
    \centering
    \includegraphics[width=\linewidth]{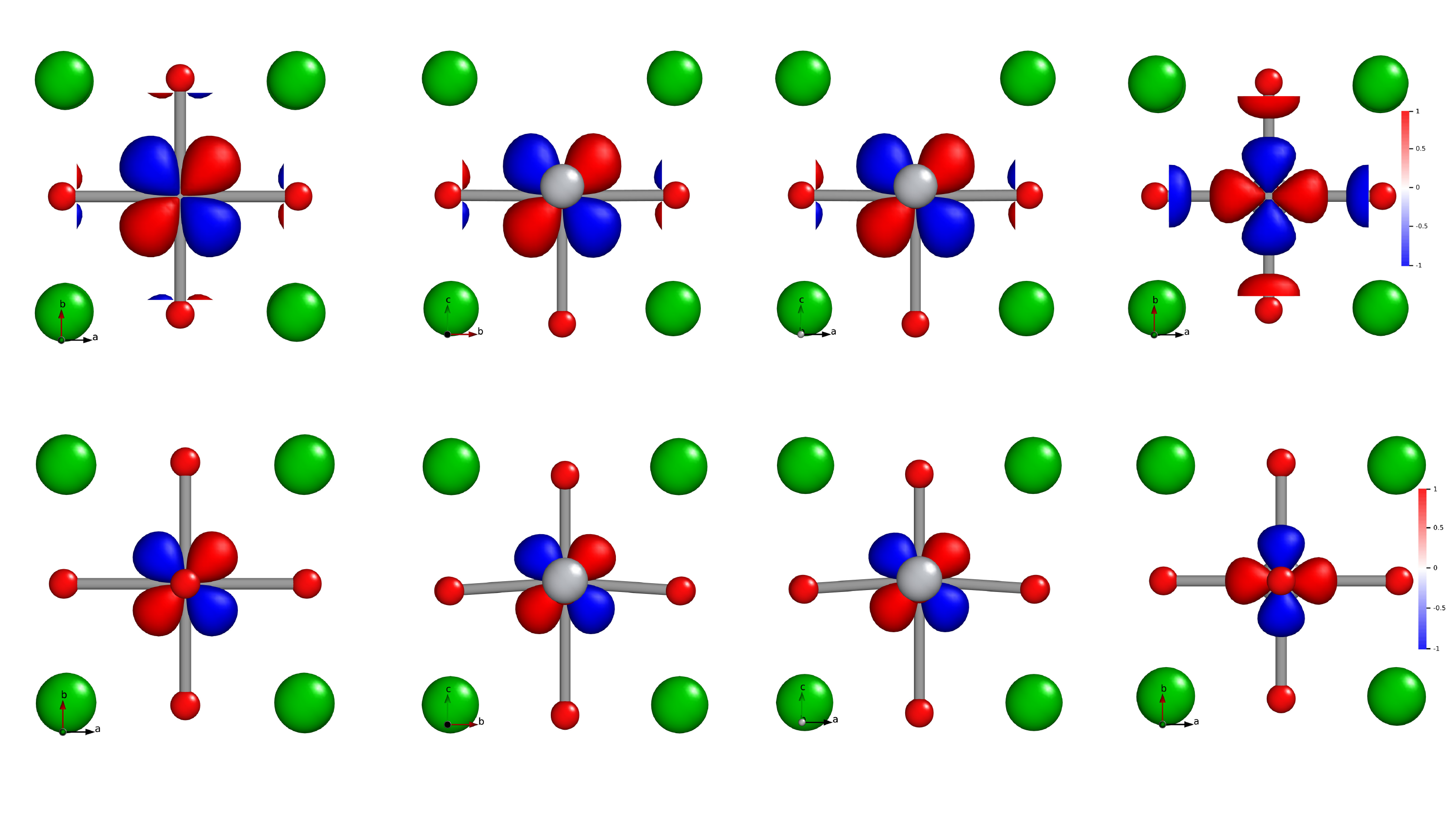}
    \caption{Wannier orbitals in order from left to right: $d_{xy}$, $d_{yz}$,
    $d_{xz}$, $d_{x^2-y^2}$, for $\mathrm{Fe_{Ti}V_O}$ (first row) and
    $\mathrm{Fe_{Ti}}$ (second row).}
    \label{fig:wannier_orb}
\end{figure}


\begin{figure}[H]
\centering
\includegraphics[width=0.6\textwidth]{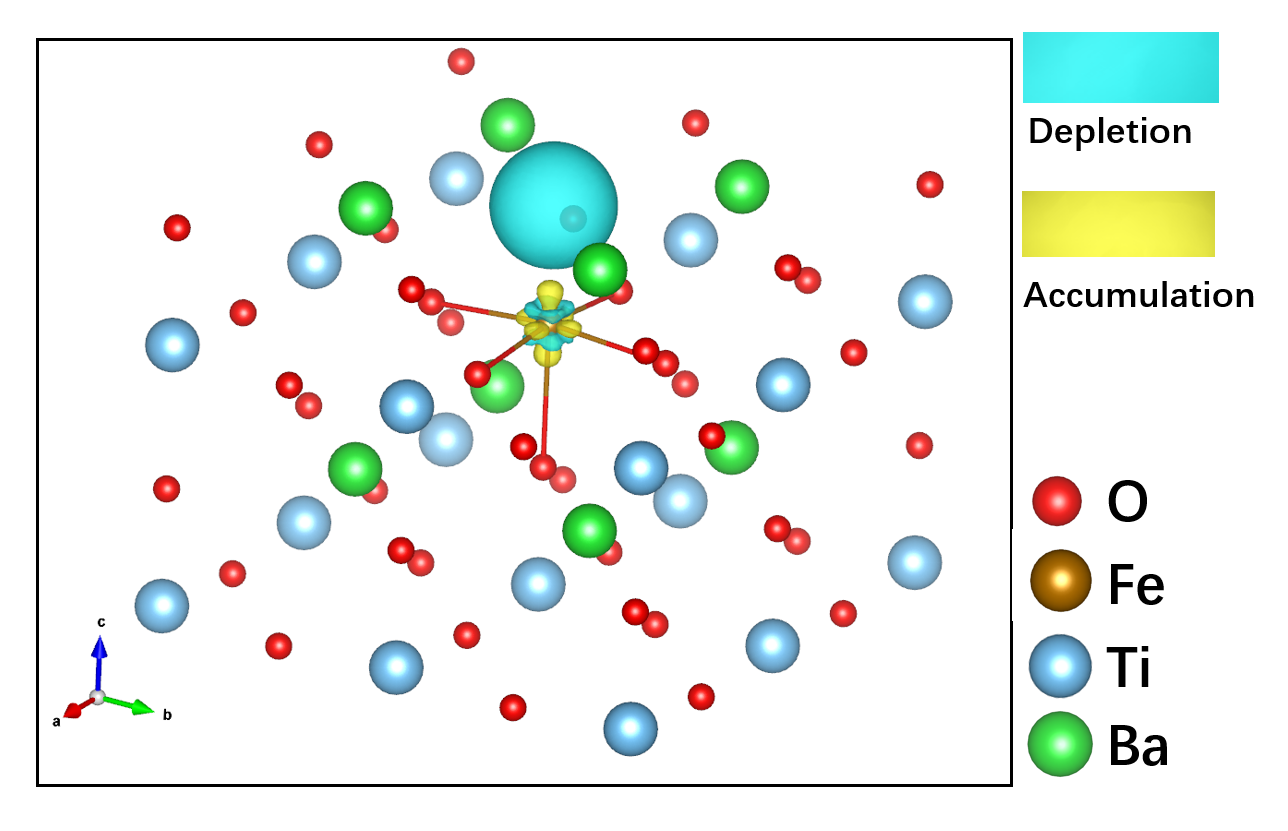}
\caption{Isosurface of the DMFT charge density difference
$\Delta \rho = \rho(\text{Fe}_{\mathrm{Ti}}\text{V}_{\mathrm{O}})
- \rho(\text{Fe}_{\mathrm{Ti}})$ plotted at 0.05\,$e$/\AA$^3$. Yellow and
cyan regions indicate electron accumulation and depletion, respectively.}
\label{fig:charge_density}
\end{figure}


\begin{figure}[H]
\centering
\includegraphics[width=0.9\textwidth]{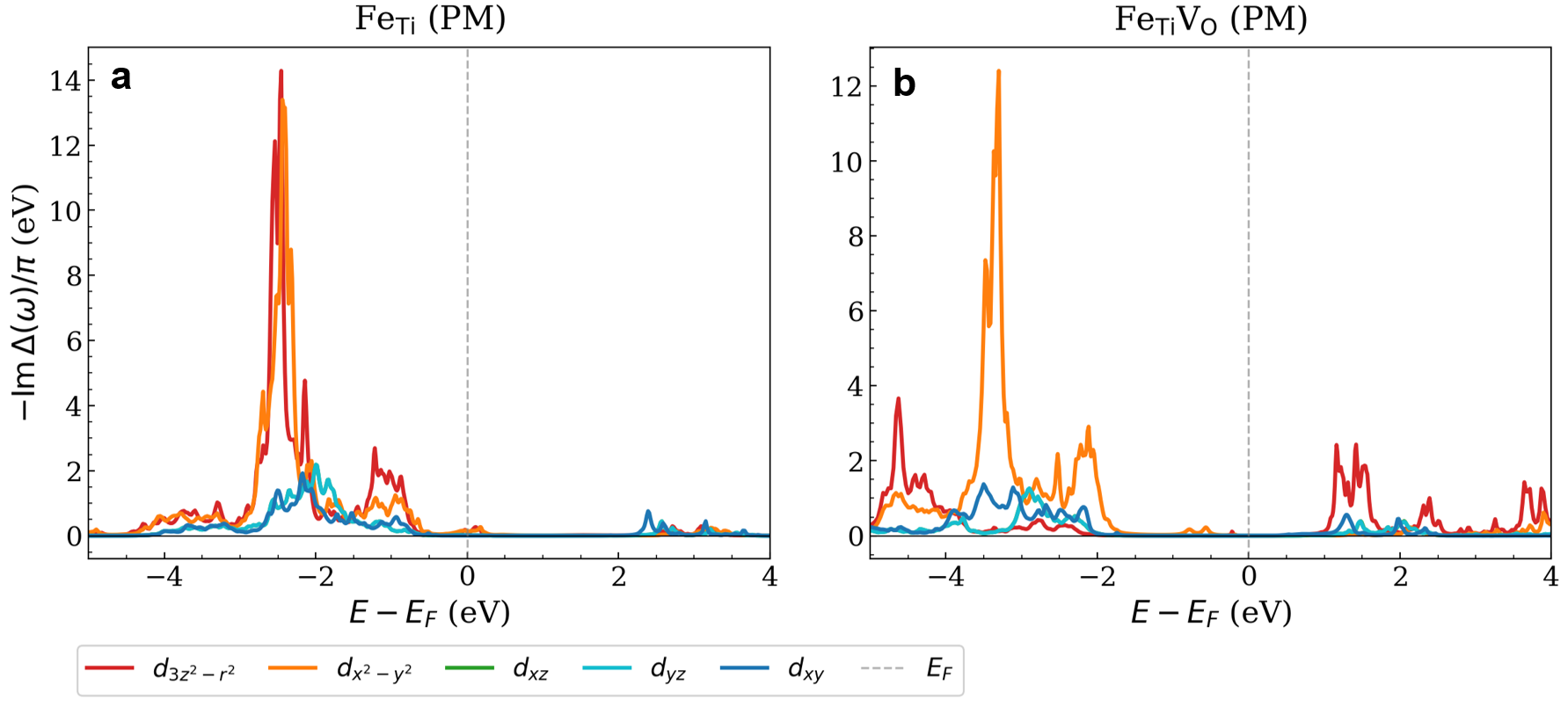}
\caption{Orbital-resolved Fe--O hybridization function
($-\mathrm{Im}\,\Delta(\omega)/\pi$) from paramagnetic DFT+DMFT for
(a)~Fe$_{\mathrm{Ti}}$ and (b)~Fe$_{\mathrm{Ti}}$V$_{\mathrm{O}}$.}
\label{fig:PM_hybridization}
\end{figure}


\begin{figure}[H]
    \centering
    \includegraphics[width=\linewidth]{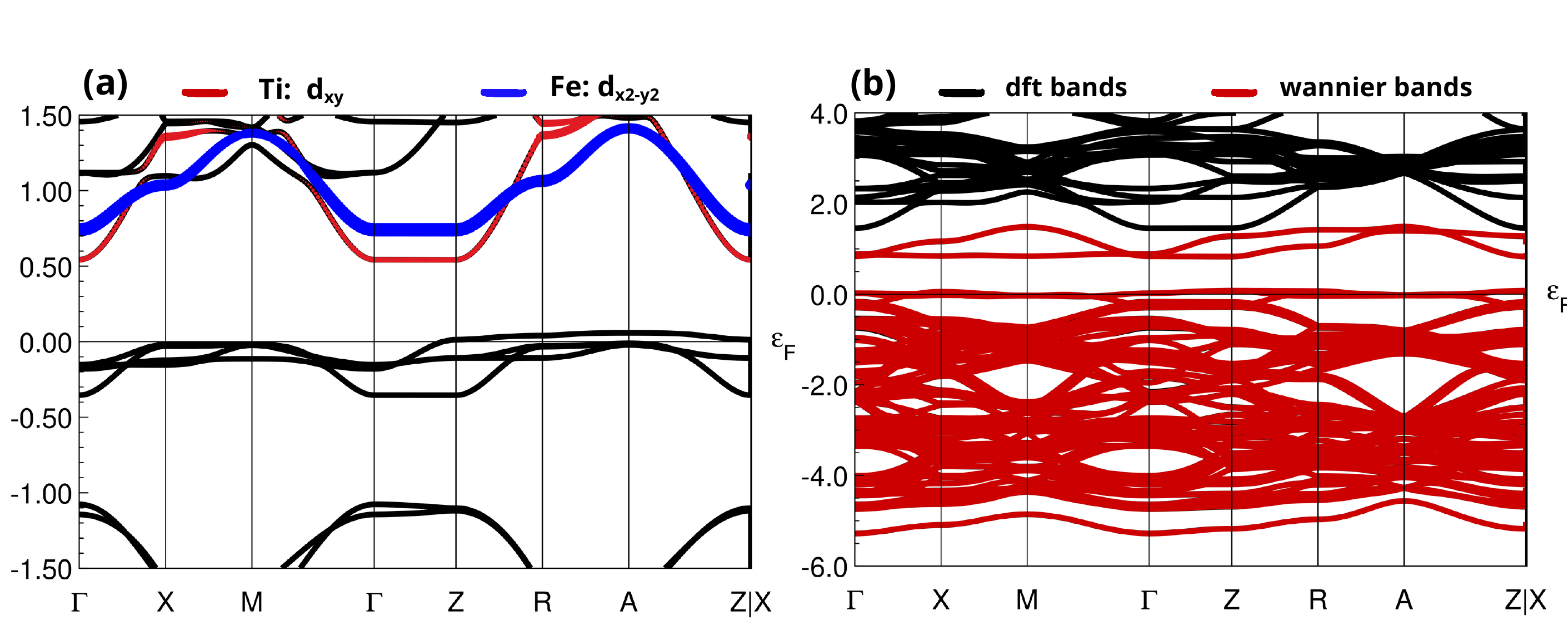}
    \caption{(a) Fat-band representation of Ti $d_{xy}$ and Fe $d_{x^2-y^2}$
    orbitals in Fe$_{\mathrm{Ti}}$V$_O$, highlighting their strong hybridization
    and the resulting entanglement that complicates the Wannier construction.
    (b) Wannier-interpolated band structure for the combined Fe $3d$ and O $2p$
    manifold (red) compared to the DFT bands (black) for the Fe$_{\mathrm{Ti}}$
    system, showing an accurate reproduction of the electronic structure.}
    \label{fig:highlighted_bands}
\end{figure}


\begin{figure}[H]
    \centering
    \includegraphics[width=0.6\linewidth]{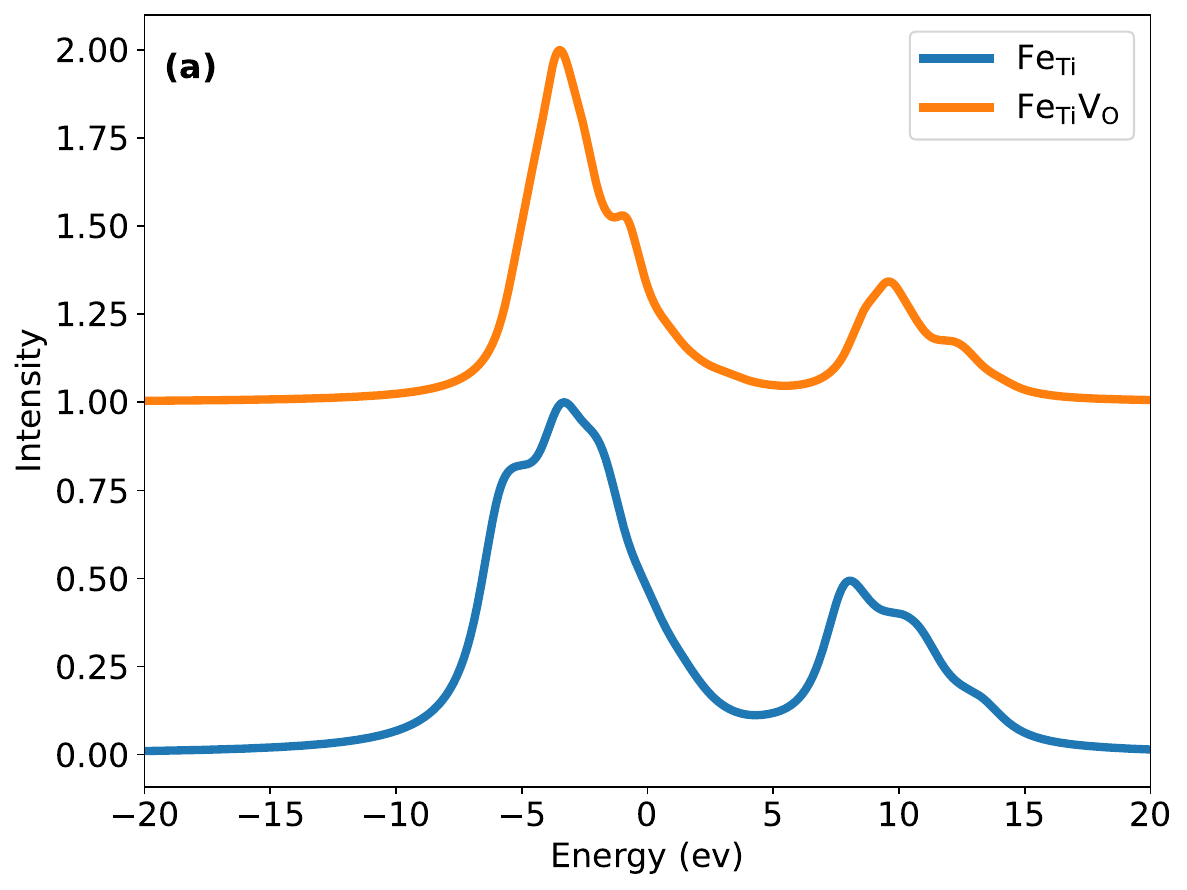}
    \caption{Calculated X-ray absorption spectra for $\mathrm{Fe_{Ti}}$ and
    $\mathrm{Fe_{Ti}V_O}$.}
    \label{fig:xas}
\end{figure}

\bibliographystyle{unsrt}
\bibliography{ref}